\documentclass[journal]{IEEEtran}

\newfont{\bb}{msbm10 scaled 1100}

\def\tr{\mathrm{tr}}

\def\diag{\mathrm{diag}}
\usepackage{graphicx}
\usepackage{algorithm,algorithmic}
\usepackage{cite}
\usepackage{psfrag}
\usepackage{url}
\usepackage{amsmath}
\usepackage{amsfonts,amssymb,amsmath}
\usepackage{mathrsfs}
\usepackage{amsthm}
\usepackage{array}

\usepackage{color}

\newtheorem{theorem}{Theorem}
\newtheorem{problem}{Problem}
\newtheorem{definition}{Definition}
\newtheorem{lemma}{Lemma}

\newtheorem{corollary}{Corollary}
\newtheorem{example}{Example}
\newtheorem{remark}{Remark}

\newcommand{\BW}{B}
\newcommand{\CU}{U}
\newcommand{\PP}{\rho}

\newcommand{\vect}[1]{\mathbf{#1}}

\newcommand{\maximize}[1]{{\underset{{#1}}{\mathrm{maximize}}}}

\begin{document}
\title{\huge{Optimal Design of Energy-Efficient Multi-User MIMO Systems: Is Massive MIMO the Answer?}}
\author{Emil Bj{\"o}rnson, \emph{Member, IEEE}, Luca Sanguinetti, \emph{Member, IEEE}, Jakob Hoydis, \emph{Member, IEEE}, and \\
M{\'e}rouane Debbah, \emph{Fellow, IEEE}
\thanks{E. Bj\"ornson is with the Department of Electrical Engineering (ISY), Link\"{o}ping University, Link\"{o}ping, Sweden (emil.bjornson@liu.se), and was previously at the KTH Royal Institute of Technology, Stockholm, Sweden, and at Sup\'elec, Gif-sur-Yvette, France.
\newline \indent L. Sanguinetti is with the University of Pisa, Dipartimento di Ingegneria dell'Informazione, Pisa, Italy (luca.sanguinetti@iet.unipi.it) and is also with the Large Systems and Networks Group (LANEAS), CentraleSup\'elec, Gif-sur-Yvette, France.
\newline \indent J. Hoydis was with Bell Laboratories, Alcatel-Lucent, Germany. He is now with Spraed SAS, Orsay, France (email: jakob.hoydis@gmail.com).
\newline \indent M. Debbah is with the Large Systems and Networks Group (LANEAS), CentraleSup\'elec, Gif-sur-Yvette, France (merouane.debbah@supelec.fr).
\newline \indent E.~Bj\"ornson was funded by ELLIIT and an International Postdoc Grant from the Swedish Research Council. L.~Sanguinetti was funded by the People Programme (Marie Curie Actions) FP7 PIEF-GA-2012-330731 Dense4Green and also by the FP7 Network of Excellence in Wireless COMmunications NEWCOM\# (Grant agreement no. 318306). This research has been supported by the ERC Starting Grant 305123 MORE (Advanced Mathematical Tools for Complex Network Engineering). This research was supported by the French p\^ole de comp\'etitivit\'e SYSTEM@TIC within the project 4G in Vitro.
\newline \indent Parts of the material in this paper were presented at the IEEE Wireless Communications and Networking Conference (WCNC) that took place in Istanbul, Turkey, April 6--9, 2014.
}}

\maketitle

\begin{abstract}Assume that a multi-user multiple-input multiple-output (MIMO) system is designed from scratch to uniformly cover a given area with maximal energy efficiency (EE). What are the optimal number of antennas, active users, and transmit power? The aim of this paper is to answer this fundamental question. We consider jointly the uplink and downlink with different processing schemes at the base station and propose a new realistic power consumption model that reveals how the above parameters affect the EE. Closed-form expressions for the EE-optimal value of each parameter, when the other two are fixed, are provided for zero-forcing (ZF) processing in single-cell scenarios. These expressions prove how the parameters interact. For example, in sharp contrast to common belief, the transmit power is found to \emph{increase} (not to decrease) with the number of antennas. This implies that energy-efficient systems can operate in high signal-to-noise ratio regimes in which interference-suppressing signal processing is mandatory. Numerical and analytical results show that the maximal EE is achieved by a massive MIMO setup wherein hundreds of antennas are deployed to serve a relatively large number of users using ZF processing. The numerical results show the same behavior under imperfect channel state information and in symmetric multi-cell scenarios.
\end{abstract}
\section{Introduction} \label{section:introduction}

The power consumption of the communication technology industry and the corresponding energy-related pollution are becoming major societal and economical concerns \cite{EARTH_D23}.
This has stimulated academia and industry to an intense activity in the new research area of \emph{green cellular networks} \cite{Chen2011a}, recently spurred by the SMART 2020 report \cite{Smart_2020} and the GreenTouch consortium \cite{GreenTouch}. The ultimate goal is to design new innovative network architectures and technologies needed to meet the explosive growth in cellular data demand without increasing the power consumption.

Along this line, in this paper we aim at jointly designing the uplink and downlink of a multi-user MIMO system for optimal energy efficiency (EE). In particular, we aim at bringing new insights on how the number $M$ of antennas at the base station (BS), the number $K$ of active user equipments (UEs), and the transmit power must be chosen in order to uniformly cover a given area with maximal EE. The EE is defined as the number of bits transferred per Joule of energy and it is affected by many factors such as (just to name a few) network architecture, transmission protocol, spectral efficiency, radiated transmit power, and circuit power consumption \cite{EARTH_D23,Smart_2020,Chen2011a,GreenTouch,Tombaz2011a}.

As discussed in \cite{Tombaz2011a}, an accurate modeling of the total power consumption is of primary importance to obtain reliable guidelines for EE optimization of $M$ and $K$. To see how this comes about, assume (as usually done in the related literature) that the total power consumption is computed as the sum of the radiated transmit power and a constant quantity accounting for the circuit power consumption \cite{EARTH_D23}. Although widely used, this model might be very misleading. In fact, it can lead to an unbounded EE if used to design systems wherein $M$ can be very large because the user rates grow unboundedly as $M \rightarrow \infty$ \cite{Bjornson2014a}. Achieving infinite EE is obviously impossible and holds true simply because the model does not take into account that the power consumed by digital signal processing and analog circuits (for radio-frequency (RF) and baseband processing) grows with $M$ and $K$. This means that its contribution can be taken as a constant only in multi-user MIMO systems where $M$ and $K$ take relatively small values, while its variability plays a key role in the so-called \emph{massive MIMO} (or large-scale MIMO) systems in which $M,K \gg 1$ and all the BS antennas are processed coherently \cite{Bjornson2014a,Marzetta2010a,Rusek2013a,Hoydis2013a,Ngo2013a}. We stress that the original massive MIMO definition in \cite{Marzetta2010a} also assumed $\frac{M}{K} \gg 1$, while we consider the more general definition from \cite{Rusek2013a} and \cite{Hoydis2013a} where $\frac{M}{K}$ can also be a small constant.

The way that the number of antennas $M$ impacts the EE has been recently investigated in \cite{Miao2013a,Hu2014a,Bjornson2013e,Ha2013a,Yang2013a,Mohammed2014a}. In particular, in \cite{Miao2013a} the author focused on the power allocation problem in the uplink of multi-user MIMO systems and showed that the EE is maximized when specific UEs are switched off. The uplink was  studied also in \cite{Hu2014a}, where the EE was shown to be a concave function of $M$ and the UE rates. The downlink was studied in \cite{Bjornson2013e,Ha2013a,Yang2013a}, whereof \cite{Bjornson2013e} and \cite{Ha2013a} showed that the EE is a concave function of $M$ while a similar result was shown for $K$ in \cite{Yang2013a}.
Unfortunately, the system parameters were optimized by means of simulations that (although useful) do not provide a complete picture of how the EE is affected by the different system parameters. The concurrent work \cite{Mohammed2014a} derives the optimal $M$ and $K$ for a given uplink sum rate, but the necessary overhead signaling for channel acquisition is ignored thereby leading to unrealistic results where it is beneficial to let $K$ grow very large, or even go to infinity.

The main purpose of this paper is to provide insights on how $M$, $K$, and the transmit power affect the total EE of a multi-user MIMO system for different linear processing schemes at the BS. The most common precoding and receive combining are considered: zero-forcing (ZF), maximum ratio transmission/combining (MRT/MRC), and minimum mean squared error (MMSE) processing \cite{Bjornson2013d}. A new refined model of the total power consumption is proposed to emphasize that the real power actually scales faster than linear with $M$ and $K$ (in sharp contrast with most existing models). Then, we concentrate on ZF processing in single-cell systems and make use of the new model for deriving closed-form EE-optimal values of each of the three system parameters, when the other two are fixed. These expressions provide valuable design insights on the interplay between system parameters, propagation environment, and different components of the power consumption model. While analytic results are given only for ZF with perfect channel state information (CSI), numerical results are provided for all the investigated schemes with perfect CSI, for ZF with imperfect CSI, and in a multi-cell scenario. Our results reveal that (a) a system with $100$-$200$ BS antennas is the right way to go if we want to be energy efficient; (b) we should use these antennas to serve a number of UEs of the same order of magnitude; (c) the transmit power should increase with the number of BS antennas since the circuit power increases; (d) ZF processing provides the highest EE due to active interference-suppression at affordable complexity.  These are highly relevant results that prove that massive MIMO is the way to achieve high EE (tens of Mbit/Joule) in future cellular networks.

The remainder of this paper is organized as follows.\footnote{The following notation is used throughout the paper. The notation $\mathbb{E}_{\mathbf{z}}\{\cdot\}$ indicates that the expectation is computed with respect to $\mathbf{z}$, whereas $||\cdot||$ and $|\cdot|$ stand for the Euclidean norm and absolute value, respectively. We let $\mathbf{I}_K$ denote the $K \times K$ identity matrix, whereas ${\bf{1}}_K$ and ${\bf{0}}_K$ are the $K$-dimensional unit and null column vectors, respectively.  We use $\mathcal{CN}(\cdot,\cdot)$ to denote a multi-variate circularly-symmetric complex Gaussian distribution. We use $e$ to indicate the natural number whereas $\ln(x)$ and $\log(x)$ denote the logarithm of $x$ to base $e$ and $2$, respectively.} In Section \ref{signalmodel}, we introduce the system model for both uplink and downlink transmissions with different linear processing schemes. The EE maximization problem is formulated in Section \ref{problem_statement} whereas the circuit power consumption model is described in Section \ref{circuit_power_model}. All this is then used in Section \ref{optimization_ZF} to compute closed-form expressions for the optimal number of UEs, number of BS antennas, and transmit power under the assumption of ZF processing. This analysis is then extended to the imperfect CSI case and to symmetric multi-cell scenarios in Section \ref{section:multicell}.
In Section \ref{numerical_results}, numerical results are used to validate the theoretical analysis and make comparisons among different processing schemes.
Finally, the major conclusions and implications are drawn in Section \ref{sec:conclusion}.

\section{System and Signal Model}\label{signalmodel}

\begin{figure}
\begin{center}
\includegraphics[width=\columnwidth]{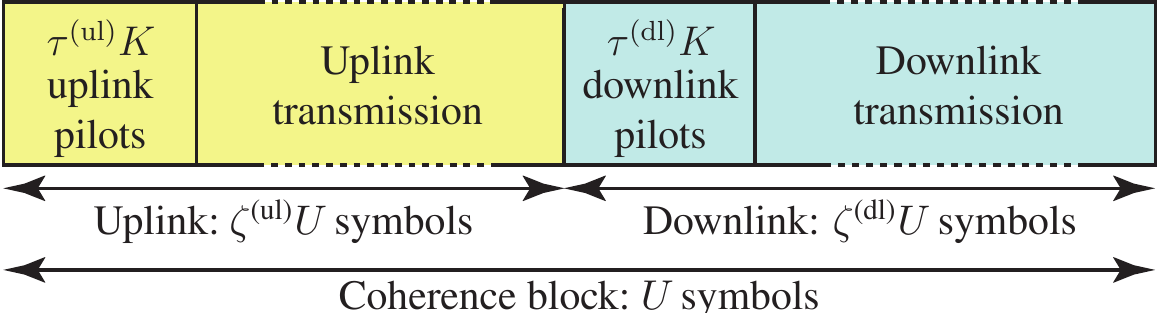}
\end{center}\vspace{-0.3cm}
\caption{Illustration of the TDD protocol, where $\zeta^{\rm{(ul)}}$ and $\zeta^{\rm{(dl)}}$ are the fractions of UL and DL transmission, respectively.} \label{figure-tdd_operation}
\end{figure}

We consider the uplink and downlink of a single-cell multi-user MIMO system operating over a bandwidth of $\BW$ Hz. The BS uses a co-located array with $M$ antennas to communicate with $K$ single-antenna UEs that are selected in round-robin fashion from a large set of UEs within the coverage area.
We consider block flat-fading channels where $\BW_{\rm{C}}$ (in Hz) is the coherence bandwidth and $T_{\rm{C}}$ (in seconds) is the coherence time. Hence, the channels are static within time-frequency coherence blocks of $\CU  = \BW_{\rm{C}} T_{\rm{C}}$ symbols. We assume that the BS and UEs are perfectly synchronized and operate according to the time-division duplex (TDD) protocol shown in Fig.~\ref{figure-tdd_operation}. The fixed ratios of uplink and downlink transmission are denoted by $\zeta^{\rm{(ul)}}$ and $\zeta^{\rm{(dl)}}$, respectively, with $\zeta^{\rm{(ul)}}+\zeta^{\rm{(dl)}}=1$. As seen from Fig.~\ref{figure-tdd_operation}, uplink transmission takes place first and consists of $\CU\zeta^{\rm{(ul)}}$ symbols. The subsequent downlink transmission consists of $\CU\zeta^{\rm{(dl)}}$ symbols. The pilot signaling occupies $\tau^{\rm{(ul)}} K$ symbols in the uplink and $\tau^{\rm{(dl)}} K$ in the downlink, where $\tau^{\rm{(ul)}},\tau^{\rm{(dl)}} \geq 1$ to enable orthogonal pilot sequences among the UEs \cite{Hoydis2013a,Ngo2013a,Bjornson2014a}. The uplink pilots enable the BS to estimate the UE channels. Since the TDD protocol is matched to the coherence blocks, the uplink and downlink channels are considered reciprocal\footnote{The physical channels are reciprocal within a coherence block, but efficient calibration schemes are needed to compensate for any possible amplitude and phase difference between the transmit and receive RF chains; we refer the reader to \cite{Zetterberg2011a} and \cite{Shepard2012a} for state-of-the-art calibration schemes.} and the BS can make use of uplink estimates for both reception and downlink transmission. TDD protocols basically require $M$ and $K$ to be the same in the uplink and downlink. The downlink pilots let each UE estimate its effective channel and interference variance with the current precoding.

\enlargethispage{4mm}

\begin{figure}
\begin{center}
\includegraphics[width=0.9\columnwidth]{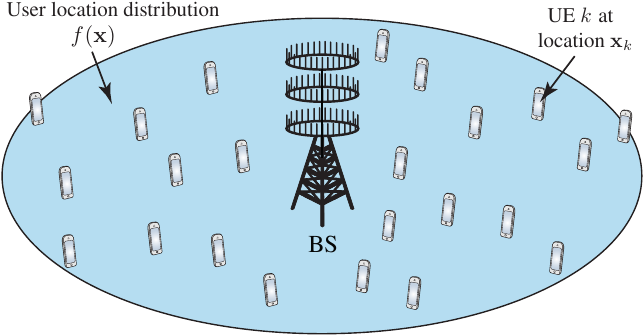}
\end{center}
\vspace{-0.3cm}
\caption{Illustration of a generic multi-user MIMO scenario: A BS with $M$ omnidirectional antennas communicates with $K$ single-antenna UEs in the uplink and downlink. The user locations are selected from an arbitrary random user distribution $f(\vect{x})$.} \label{figure_scenario}
\end{figure}

The physical location of UE $k$ is denoted by $\mathbf{x}_k \in \mathbb{R}^2$ (in meters) and is computed with respect to the BS (assumed to be located in the origin). For analytic tractability, we consider only non-line-of-sight propagation. The function $l(\cdot): \mathbb{R}^2 \rightarrow \mathbb{R}$ describes the large-scale channel fading at different user locations; that is, $l(\mathbf{x}_k )$ is the average channel attenuation\footnote{It is also known as \emph{channel gain}, but since we deal with EE we stress that channels attenuate rather than amplify signals.} due to path-loss, scattering, and shadowing at location $\mathbf{x}_k$. Since the UEs are selected in a round-robin fashion, the user locations can be treated as random variables from a user distribution $f(\vect{x})$ implicitly defining the shape and user density of the coverage area (see Fig.~\ref{figure_scenario}). The large-scale fading between a UE and the BS is assumed to be the same for all BS antennas. This is reasonable since the distances between UEs and the BS are much larger than the distance between the antennas. Since the forthcoming analysis does not depend on a particular choice of $l(\cdot)$ and user distribution, we keep it generic. The following symmetric example is used for simulations.

\begin{example} \label{example:simulation-scenario}
Suppose the UEs are uniformly distributed in a circular cell with radius $d_{\max}$ and minimum distance $d_{\min}$. This user distribution is described by the density function
\begin{equation}
f(\vect{x}) = \begin{cases} \frac{1}{\pi (d_{\max}^2 - d_{\min}^2 )} & d_{\min} \leq \| \vect{x} \| \leq d_{\max}, \\ 0 & \text{otherwise}. \end{cases}
\end{equation}
Moreover, let the large-scale fading be dominated by path-loss. This is often modeled as
\begin{equation}
l(\vect{x}) = \frac{\bar d}{\| \vect{x} \|^{\kappa}} \quad \quad  \text{for} \,\,\,\,\, \|\vect{x}\|\geq d_{\min}
\end{equation}
where $\kappa \geq 2$ is the path-loss exponent and the constant $\bar d>0$ regulates the channel attenuation at distance $d_{\min}$ \cite{LTE2010b}. The average inverse channel attenuation, $\mathbb{E}_\mathbf{x} \{  (l(\mathbf{x}))^{-1}\}$ plays a key role in all subsequent discussions. In this example, simple integration (using polar coordinates) shows that
\begin{equation}\label{S_x}
\mathbb{E}_\mathbf{x} \left\{  \big(l(\mathbf{x}) \big)^{-1}\right\} = \frac{d_{\max}^{\kappa+2} - d_{\min}^{\kappa+2}}{\bar d ( 1 + \frac{\kappa}{2})(d_{\max}^2 - d_{\min}^2)}.
\end{equation}
\end{example}

\subsection{Channel Model and Linear Processing}

The $M$ antennas at the BS are adequately spaced apart such that the channel components between the BS antennas and the single-antenna UEs are uncorrelated. The channel vector $\mathbf{h}_{k} = [{h}_{k,1},{h}_{k,2},\ldots,{h}_{k,M}]^T \in \mathbb{C}^{M\times 1}$ has entries $\{h_{k,n}\}$ that describe the instantaneous propagation channel between the $n$th antenna at the BS and the $k$th UE. We assume a Rayleigh small-scale fading distribution such that $\mathbf{h}_{k} \sim \mathcal{CN} \big( \vect{0}_M, l(\mathbf{x}_k )\vect{I}_M  \big)$, which is a valid model for both small and large arrays \cite{Gao2015a}. Linear processing is used for uplink data detection and downlink data precoding.
For analytic tractability, we assume that the BS is able to acquire perfect CSI from the uplink pilots; the imperfect CSI case is considered in Section \ref{section:multicell}. We denote the uplink linear receive combining matrix by $\mathbf{G} = [\vect{g}_1, \vect{g}_2, \ldots, \vect{g}_K] \in \mathbb{C}^{M\times K}$ with the column $\vect{g}_k$ being assigned to the $k$th UE.  We consider MRC, ZF, and MMSE for uplink detection, which gives
\begin{align} \label{eq:def-G}
\mathbf{G} = \left\{ {\begin{array}{*{20}{l}}
\mathbf{H}& \text{for MRC,}\\
\mathbf{H} \left(\mathbf{H}^H\mathbf{H}\right)^{-1} & \text{for ZF,}\\
 \left(\mathbf{H} \vect{P}^{(\rm{ul})} \mathbf{H}^H + \sigma^2 \mathbf{I}_M\right)^{-1} \mathbf{H} & \text{for MMSE,}\\
\end{array}} \right.
\end{align}
where $\mathbf{H} = [\mathbf{h}_1,\mathbf{h}_2,\ldots,\mathbf{h}_K]$ contains all the user channels, $\sigma^2$ denotes the noise variance (in Joule/symbol), $ \vect{P}^{(\rm{ul})} = \diag({p_1^{(\rm{ul})}},{p_2^{(\rm{ul})}},\ldots,{p_K^{(\rm{ul})}})$, and the design parameter ${p_i^{(\rm{ul})}} \geq 0$ is the transmitted uplink power of UE $i$ (in Joule/symbol) for $i=1,2,\ldots,K$. Similarly, we consider MRT, ZF, and transmit-MMSE as precoding schemes for downlink transmissions \cite{Bjornson2013d}. Denoting by $\mathbf{V} = [\vect{v}_1, \vect{v}_2, \ldots, \vect{v}_K] \in \mathbb{C}^{M\times K}$ the precoding matrix, we have that
\begin{align} \label{eq:def-V}
\mathbf{V} = \left\{ {\begin{array}{*{20}{l}}
\mathbf{H}& \text{for MRT,}\\
\mathbf{H}\left(\mathbf{H}^H \mathbf{H}\right)^{-1} & \text{for ZF,}\\
 \left(\mathbf{H} \vect{P}^{(\rm{ul})} \mathbf{H}^H + \sigma^2 \mathbf{I}_M\right)^{-1} \mathbf{H} & \text{for MMSE.}\\
\end{array}} \right.
\end{align}
It is natural to set $\vect{V} = \vect{G}$, since it reduces the computational complexity, but it is not necessary.

While conventional systems have large disparity between peak and average rates, we aim at designing the system so as to guarantee a uniform \emph{gross} rate $\bar{R}$ (in bit/second) for any active UE, whereof $\zeta^{\rm{(ul)}} \bar{R}$ is the uplink rate and $\zeta^{\rm{(dl)}} \bar{R}$ is the downlink rate. As detailed below, this is achieved by combining the linear processing with proper power allocation.

\subsection{Uplink}
Under the assumptions of Gaussian codebooks, linear processing, and perfect CSI \cite{Hoydis2013a}, the achievable uplink rate (in bit/second) of the $k$th UE is
\begin{align} \label{eq:rate-expression_ul}
R_k^{(\rm{ul})} =  \zeta^{(\rm{ul})}  \left( 1-\frac{\tau^{\rm{(ul)}} K}{\CU \zeta^{(\rm{ul})}} \right)\bar R_k^{(\rm{ul})}
\end{align}
where the pre-log factor $\left( 1-\frac{\tau^{\rm{(ul)}} K}{\CU \zeta^{(\rm{ul})}} \right)$ accounts for pilot overhead and $\zeta^{(\rm{ul})}$ is the fraction of uplink transmission. In addition,
\begin{align}\label{rate_ul}
\bar {R}_k^{(\rm{ul})} = \BW \log \Bigg( 1 + \frac{ p_k^{(\rm{ul})}|\vect{g}_k^H \vect{h}_k|^2 }{ \sum\limits_{\ell =1,\ell \neq k}^{K}  p_\ell^{(\rm{ul})}|\vect{g}_k^H \vect{h}_{\ell}|^2 + \sigma^2\left\|\vect{g}_k\right\|^2}  \Bigg)
\end{align}
is the uplink \emph{gross} rate (in bit/second) from the $k$th UE, where ``gross'' refers to that overhead factors are not included.
As mentioned above, we aim at providing the same gross rate $\bar R_k^{\rm{(ul)}} = \bar{R}$ for $k=1,2,\ldots,K$. By utilizing a technique from \cite{Pillai2005a}, this equal-rate condition is met if and only if the uplink power allocation vector $\mathbf{p}^{(\rm{ul})} = [{p_1^{(\rm{ul})}}, {p_2^{(\rm{ul})}},\ldots,{p_K^{(\rm{ul})}}]^T$ is such that
\begin{align}\label{MRC.1}
\mathbf{p}^{(\rm{ul})} = \sigma^2 (\mathbf{D}^{{(\rm{ul})}})^{-1} \mathbf{1}_{K}
\end{align}
where the $(k,\ell)$th element of $\mathbf{D}^{(\rm{ul})} \in \mathbb C^{K\times K}$ is
\begin{align}\label{D_UL}
\left[\mathbf{D}^{(\rm{ul})}\right]_{k,\ell}= \begin{cases}
\frac{|\vect{g}_k^H \vect{h}_k|^2}{(2^{\bar{R}/\BW}-1) \|\vect{g}_k\|^2}& \text{for} \,\,\, k = \ell, \\
-\frac{|\vect{g}_k^H \vect{h}_{\ell}|^2}{\|\vect{g}_k\|^2}& \text{for} \,\,\, k \ne \ell.
\end{cases}
\end{align}
The power allocation in \eqref{MRC.1} is computed directly for MRC and ZF detection, while it is a fixed-point equation for MMSE detection since also $\vect{G}$ depends on the power allocation \cite{Wiesel2006a}.

The average uplink \emph{PA power} (in Watt) is defined as the power consumed by the power amplifiers (PAs), which includes radiated transmit power and PA dissipation.
By using \eqref{MRC.1} it is found to be\footnote{We assume that the average transmit power is the same in both phases of the uplink slot, but it might be fixed during pilot signaling and time-varying for data transmission; see Section \ref{section:multicell}. UE $k$ computes its power ${p_k^{(\rm{ul})}}$ in the previous downlink slot.}
\begin{align} \label{power_UL.11}
P_\mathrm{TX}^{(\rm{ul})} =
\frac{\BW\zeta^{(\rm{ul})}}{\eta^{(\rm{ul})}} \mathbb{E}\{ \mathbf{1}_{K}^T  \mathbf{p}^{(\rm{ul})} \}   = \sigma^2 \frac{\BW \zeta^{(\rm{ul})}}{\eta^{(\rm{ul})}} \mathbb{E} \left\{ \mathbf{1}_{K}^T  (\mathbf{D}^{(\rm{ul})})^{-1} \mathbf{1}_{K} \right\}
\end{align}
where $0< \eta^{(\rm{ul})} \leq 1$ is the PA efficiency at the UEs.

Observe that it might happen that $\bar{R}$ cannot be supported for any transmit powers. In such a case, computing $\mathbf{p}^{(\rm{ul})}$ in \eqref{MRC.1} would lead to some negative powers. However, this can easily be detected and avoided by computing the spectral radius of $\mathbf{D}^{{(\rm{ul})}}$ \cite{Pillai2005a}. Moreover, it only happens in interference-limited cases; thus, it is not an issue when ZF is employed (under perfect CSI). In these circumstances, $P_\mathrm{TX}^{(\rm{ul})}$ in \eqref{power_UL.11} can be computed in closed form as stated in the following.

\begin{lemma}\label{UL-ZF_TX}
If a ZF detector is employed with $M \ge K+1$, we can without loss of generality parameterize the gross rate as
\begin{align}\label{constant_rate}
\bar{R} = B \log \left( 1 + \PP\left(M-K\right)
  \right)
\end{align}
where $\PP$ is a design parameter that is proportional to the received signal-to-interference-and-noise ratio (SINR). Using this parameterization, the PA power $P_\mathrm{TX}^{(\rm{ul-ZF})}$ required to guarantee each UE the gross rate in \eqref{constant_rate} is 
\begin{align}\label{power_UL_ZF}
P_\mathrm{TX}^{(\rm{ul-ZF})} =  \frac{\BW \zeta^{(\rm{ul})}}{\eta^{(\rm{ul})}} \sigma^2 \PP \mathcal{S}_{\mathbf{x}} K
\end{align}
where $\mathcal{S}_{\mathbf{x}} = \mathbb{E}_\mathbf{x} \left\{{ (l(\mathbf{x})})^{-1}\right\}$
accounts for user distribution and propagation environment.
\end{lemma}
\begin{IEEEproof}
This result is proved in the appendix.
\end{IEEEproof}

The gross rate in \eqref{constant_rate} is used for ZF processing in the remainder of this paper, since it gives simple PA power expressions. The parameter $\PP$ is later treated as an optimization variable.

\subsection{Downlink}

The downlink signal to the $k$th UE is assigned a transmit power of ${p}_k^{(\rm{dl})}$ (in Joule/symbol) and a normalized precoding vector ${\vect{v}_k}/{\| \vect{v}_k\|}$.
Assuming Gaussian codebooks and perfect CSI \cite{Bjornson2013d}, the achievable downlink rate (in bit/second) of the $k$th UE with linear processing is
\begin{align} \label{eq:rate-expression}
R_k ^{(\rm{dl})} =  {\zeta^{(\rm{dl})}} \left( 1-\frac{\tau^{\rm{(dl)}} K}{\CU\zeta^{(\rm{dl})}} \right) \bar R_k^{(\rm{dl})}
\end{align}
where $\left( 1-\frac{\tau^{\rm{(dl)}} K}{\CU\zeta^{(\rm{dl})}} \right)$ accounts for the downlink pilot overhead and $\bar {R}_k^{(\rm{dl})}$ is the gross rate (in bit/second) given by
\begin{align}\label{SINR_dl}
\bar {R}_k^{(\rm{dl})} = \BW \log \Bigg( 1 +\frac{ p_k^{(\rm{dl})} \frac{|\vect{h}_k^H \vect{v}_k|^2}{ \| \vect{v}_k \|^2 } }{ \sum\limits_{\ell =1,\ell \neq k}^{K} p_\ell^{(\rm{dl})} \frac{|\vect{h}_k^H \vect{v}_{\ell}|^2}{\| \vect{v}_{\ell} \|^2 } + \sigma^2}\Bigg).
\end{align}
The average PA power is defined as
\begin{align}\label{power_DL}
P_\mathrm{TX}^{(\rm{dl})} =  \frac{B \zeta^{(\rm{dl})}}{\eta^{(\rm{dl})}} \sum\limits_{k=1}^K \mathbb{E} \left\{ p_k^{(\rm{dl})} \right\}
\end{align}
where $0< \eta^{(\rm{dl})} \leq 1$ is the PA efficiency at the BS.
Imposing the equal-rate condition $\bar R_k^{(\rm{dl})} = \bar R$ for all $k$, it follows that the power allocation vector $\mathbf{p}^{(\rm{dl})} = [{p}_1^{(\rm{dl})},{p}_2^{(\rm{dl})},\ldots,{p}_K^{(\rm{dl})}]^T$  must be computed as $\mathbf{p}^{(\rm{dl})} = \sigma^2 (\mathbf{D}^{(\rm{dl})})^{-1}\mathbf{1}_{K}$ \cite{Pillai2005a},
where the $(k,\ell)$th element of $\mathbf{D}^{(\rm{dl})} \in \mathbb C^{K\times K}$ is
\begin{align}\label{DL.2}
\left[\mathbf{D}^{(\rm{dl})}\right]_{k,\ell}= \begin{cases}
 \frac{|\vect{h}_k^H \vect{v}_k|^2}{(2^{\bar{R}/B} -1)  \| \vect{v}_k \|^2} & \text{for} \,\, k = \ell, \\
-\frac{|\vect{h}_k^H \vect{v}_{\ell}|^2}{\| \vect{v}_{\ell}\|^2} & \text{for} \,\, k \ne \ell.
\end{cases}
\end{align}
Plugging $\mathbf{p}^{(\rm{dl})} = \sigma^2 (\mathbf{D}^{(\rm{dl})})^{-1}\mathbf{1}_{K}$ into \eqref{power_DL}, the average downlink PA power (in Watt) is
\begin{align}\label{DL.5}
P_\mathrm{TX}^{(\rm{dl})} = \sigma^2 \frac{\BW \zeta^{(\rm{dl})}}{\eta^{(\rm{dl})}}
\mathbb{E}\left\{  \mathbf{1}_K^T (\mathbf{D}^{{(\rm{dl})}})^{-1}\mathbf{1}_K \right\}.
\end{align}
Observe that $\mathbf{D}^{(\rm{dl})} = (\mathbf{D}^{(\rm{ul})})^T$ if the same processing scheme  is used for transmit precoding and receive combining (i.e., if $\vect{G} = \vect{V}$). In this case, the user-specific uplink/downlink transmit powers are different, but the total uplink and downlink PA powers in \eqref{power_UL.11} and \eqref{DL.5}, respectively, are the same (except for the factors ${\zeta^{(\rm{ul})}}/{\eta^{(\rm{ul})}}$ and ${\zeta^{(\rm{dl})}}/{\eta^{(\rm{dl})}}$). This is a consequence of the well-known uplink-downlink duality \cite{Boche2002a}.

Similar to the uplink, the following result can be proved for ZF in the downlink.
\begin{lemma}\label{DL-ZF_TX}
If ZF precoding is used with $M \ge K+1$, then the average downlink PA power $P_\mathrm{TX}^{(\rm{dl-ZF})}$ required to serve each UE with a gross rate equal to $\bar{R}$ in \eqref{constant_rate} is
\begin{align} \label{power_DL.1}
P_\mathrm{TX}^{(\rm{dl - ZF})} = \frac{\BW \zeta^{(\rm{dl})}}{\eta^{(\rm{dl})}} \sigma^2 \PP \mathcal{S}_\mathbf{x} K
\end{align}
where $\mathcal{S}_\mathbf{x}$ is the propagation environment parameter defined in Lemma \ref{UL-ZF_TX}.
\end{lemma}
\begin{IEEEproof}
This result is proved in the appendix.
\end{IEEEproof}

From Lemmas \ref{UL-ZF_TX} and \ref{DL-ZF_TX}, it is seen that the average uplink and downlink PA powers sum up to
\begin{align} \label{power_DL.10}
P_\mathrm{TX}^{(\rm{ZF})}  = P_\mathrm{TX}^{(\rm{ul-ZF})} + P_\mathrm{TX}^{(\rm{dl-ZF})} =
\frac{\BW \sigma^2 \PP \mathcal{S}_\mathbf{x}}{\eta} K
\end{align}
under ZF processing, where $\eta = \Big(\frac{\zeta^{(\rm{ul})}}{\eta^{(\rm{ul})}} + \frac{\zeta^{(\rm{dl})}}{\eta^{(\rm{dl})}}\Big)^{-1}$.
\begin{remark}
A key assumption in this paper is that a uniform gross rate $\bar{R}$ is guaranteed to all UEs by means of power allocation. However, the main results are also applicable in cases with fixed power allocation. Suppose for example that the transmit power is allocated equally under ZF processing. Then, the Jensen's inequality can be used (as is done in \cite{Caire2013a}) to prove that $\bar{R}$ is a lower bound of the average gross rates $\mathbb{E}\{\bar {R}_k^{(\rm{ul})}\}$ and $\mathbb{E}\{\bar {R}_k^{(\rm{dl})}\}$ (where the expectations are taken with respect to both user locations and channel realizations).
\end{remark}

\section{Problem Statement}\label{problem_statement}

As mentioned in Section \ref{section:introduction}, the EE of a communication system is measured in bit/Joule \cite{Chen2011a} and is computed as the ratio between the average sum rate (in bit/second) and the average total power consumption $P_{\mathrm{T}}$ (in Watt = Joule/second). In a multi-user setting, the \emph{total} EE metric accounting for both uplink and downlink takes the following form.
\begin{definition} The total EE of the uplink and downlink is
\begin{equation} \label{eq:EE-metric}
\mathrm{EE} =  \frac{ \sum\limits_{k=1}^K \Big(\mathbb{E} \left\{ R_k^{(\rm{ul})} \right\} + \mathbb{E} \left\{ R_k^{(\rm{dl})} \right\} \Big)}{ P_{\mathrm{TX}}^{(\rm{ul})} + P_{\mathrm{TX}}^{(\rm{dl})} + P_{\mathrm{CP}} }
\end{equation}
where $P_{\mathrm{CP}}$ accounts for the circuit power consumption.
\end{definition}
In most of the existing works, $P_{\mathrm{CP}}$ is modeled as $P_{\mathrm{CP}} = P_{\mathrm{FIX}}$
where the term $P_{\mathrm{FIX}}$ is a constant quantity accounting for the \emph{fixed} power consumption required for site-cooling, control signaling, and  load-independent power of backhaul infrastructure and baseband processors \cite{EARTH_D23}. This is not an accurate model if we want to design a good system by optimizing the number of antennas ($M$) and number of UEs ($K$); in fact, Lemmas \ref{UL-ZF_TX} and \ref{DL-ZF_TX} show that the achievable rates with ZF grow logarithmically with $M$ (for a fixed PA power). Hence, the simplified model $P_{\mathrm{CP}} = P_{\mathrm{FIX}}$ gives the impression that we can achieve an unbounded EE by adding more and more antennas. This modeling artifact comes from ignoring that each antenna at the BS requires dedicated circuits with a non-zero power consumption, and that the signal processing tasks also become increasingly complex.

In other words, an accurate modeling of $P_{\mathrm{CP}}$ is of paramount importance when dealing with the design of energy-efficient communication systems. The next section aims at providing an appropriate model for $P_{\mathrm{CP}}(M,K,\bar{R})$ as a function of the three main design parameters: the number of BS antennas ($M$), number of active UEs ($K$), and the user gross rates ($\bar R$).

Based on this model, we now formulate the main problem of this paper.

\begin{problem} \label{problem-statement}
An EE-optimal multi-user MIMO setup is achieved by solving the following optimization problem: 
\begin{equation} \label{eq:problem-statement}
\maximize{M \in \mathbb{Z}_+, \, K \in \mathbb{Z}_+, \, \bar R \geq 0} \quad \mathrm{EE} =  \frac{ \sum\limits_{k=1}^K \Big(\mathbb{E} \{ R_k^{(\rm{ul})} \} + \mathbb{E} \{ R_k^{(\rm{dl})} \} \Big)}{ P_{\mathrm{TX}}^{(\rm{ul})} + P_{\mathrm{TX}}^{(\rm{dl})} + P_{\mathrm{CP}}(M,K,\bar{R}) }.
\end{equation}
\end{problem}

This problem is solved analytically for ZF processing in Section \ref{optimization_ZF} and numerically in Section \ref{numerical_results} for other processing schemes.

\begin{remark}
Observe that prior works on EE optimization have focused on either uplink or downlink. In contrast, Problem \ref{problem-statement} is a holistic optimization in which the total EE is maximized for given fractions $\zeta^{\rm{(ul)}}$ and $\zeta^{\rm{(dl)}}$ of uplink and downlink transmissions. The optimization of the uplink or downlink only is clearly a special case in which $\zeta^{\rm{(ul)}} = 0 $ or $\zeta^{\rm{(dl)}} = 0$, respectively.
\end{remark}

\begin{remark}
Maximizing the EE in \eqref{eq:problem-statement} does not mean decreasing the total power, but to pick a good power level and use it wisely.
Section \ref{numerical_results} indicates that future networks can increase the EE by having much higher sum rates, but at the cost of also increasing the power consumption.
\end{remark}

\section{Realistic Circuit Power Consumption Model}\label{circuit_power_model}

The circuit power consumption $ P_{\mathrm{CP}}$ is the sum of the power consumed by different analog components and digital signal processing \cite{EARTH_D23}. Building on the prior works of \cite{Tombaz2011a,EARTH_D23,Yang2013a,Cui2004a,Mezghani2011a,Kumar2011a}, we propose a new refined circuit power consumption model for multi-user MIMO systems: 
\begin{align}\label{P_CP}
P_{\mathrm{CP}} = P_{\mathrm{FIX}} + P_\mathrm{TC} + P_\mathrm{CE} + P_\mathrm{C/D} + P_\mathrm{BH} + P_\mathrm{LP}
\end{align}
where the fixed power $P_{\mathrm{FIX}}$ was defined in Section \ref{problem_statement}, $P_\mathrm{TC}$ accounts for the power consumption of the transceiver chains, $P_\mathrm{CE}$ of the channel estimation process (performed once per coherence block), $P_\mathrm{C/D}$ of the channel coding and decoding units, $P_\mathrm{BH}$ of the load-dependent backhaul, and $P_\mathrm{LP}$ of the linear processing at the BS. In the following, we provide simple and realistic models for how each term in \eqref{P_CP} depends, linearly or non-linearly, on the main system parameters $(M,K,\bar{R})$. This is achieved by characterizing the hardware setup using a variety of fixed coefficients, which are kept generic in the analysis; typical values are given later in Table \ref{table_parameters_hardware}. The proposed model is inspired by \cite{Tombaz2011a,EARTH_D23,Yang2013a,Cui2004a,Mezghani2011a,Kumar2011a,Boyd2008a}, but goes beyond these prior works by modeling all the terms with realistic, and sometimes non-linear, expressions.

\subsection{Transceiver Chains}

As described in \cite{Cui2004a} and \cite{Kumar2011a}, the power consumption $P_\mathrm{TC}$ of a set of typical transmitters and receivers can be quantified as 
\begin{align}\label{P_{TC}}
P_\mathrm{TC} = M P_{\mathrm{BS}} + P_{\mathrm{SYN}} + K P_{\mathrm{UE}} \quad {\text{Watt}}
\end{align}
where $P_{\mathrm{BS}}$ is the power required to run the circuit components (such as converters, mixers, and filters) attached to each antenna at the BS and $P_{\mathrm{SYN}}$ is the power consumed by the local oscillator.\footnote{In general, a single oscillator is used for frequency synthesis at all BS antennas. This is the reason that this term is independent of $M$. If multiple oscillators are used (e.g., for distributed antenna arrays) we can easily set $P_{\mathrm{SYN}}=0$ and include the power consumption of the oscillators in $ P_{\mathrm{BS}}$ instead.} The last term $P_{\mathrm{UE}}$ accounts for the power required by all circuit components (such as amplifiers, mixer, oscillator, and filters) of each single-antenna UE.

\subsection{Channel Estimation}

All processing is carried out locally at the BS and UEs, whose computational efficiency are $L_{\rm{BS}}$ and $L_{\rm{UE}}$ arithmetic complex-valued operations per Joule (also known as flops/Watt), respectively. There are $\frac{\BW}{\CU}$ coherence blocks per second and the pilot-based CSI estimation is performed once per block. In the uplink, the BS receives the pilot signal as an $M \times \tau^{\rm{(ul)}} K$ matrix and estimates each UE's channel by multiplying with the corresponding pilot sequence of length $\tau^{\rm{(ul)}} K$ \cite{Hoydis2013a}. This a standard linear algebra operation \cite{Boyd2008a} and requires $P_\mathrm{CE}^{(\rm{ul})} = \frac{\BW}{\CU} \frac{2\tau^{\rm{(ul)}} MK^2}{L_{\rm{BS}}}$ Watt.
In the downlink, each active UE receives a pilot sequence of length $\tau^{\rm{(dl)}} K$ and processes it to acquire its effective precoded channel gain (one inner product) and
 the variance of interference plus noise (one inner product). From \cite{Boyd2008a}, we obtain  $P_\mathrm{CE}^{(\rm{dl})} = \frac{\BW}{\CU} \frac{4 \tau^{\rm{(dl)}} K^2}{L_{\rm{UE}}}$ Watt. Therefore, the total power consumption $P_\mathrm{CE} = P_\mathrm{CE}^{(\rm{ul})} + P_\mathrm{CE}^{(\rm{dl})} $ of the channel estimation process becomes
\begin{align}\label{P_CE}
P_\mathrm{CE} = \frac{\BW}{\CU} \frac{2\tau^{\rm{(ul)}} MK^2}{L_{\rm{BS}}} + \frac{\BW}{\CU} \frac{4 \tau^{\rm{(dl)}} K^2}{L_{\rm{UE}}}  \quad {\text{Watt}}.
\end{align}

\subsection{Coding and Decoding}

In the downlink, the BS applies channel coding and modulation to $K$ sequences of information symbols and each UE applies some suboptimal fixed-complexity algorithm for decoding its own sequence. The opposite is done in the uplink. The power consumption $P_{\mathrm{C/D}}$ accounting for these processes is proportional to the number of bits \cite{Mezghani2011a} and can thus be quantified as
\begin{align}\label{P_C/D}
P_{\mathrm{C/D}} =  \sum\limits_{k=1}^K \left(  \mathbb{E} \{ R_k^{(\rm{ul})} +  R_k^{(\rm{dl})} \}  \right) ( P_{\mathrm{COD}} +P_{\mathrm{DEC}}) \quad {\text{Watt}}
\end{align}
where $P_{\mathrm{COD}}$ and $P_{\mathrm{DEC}}$ are the coding and decoding powers (in Watt per bit/s), respectively. For simplicity, we assume that $P_{\mathrm{COD}}$ and $P_{\mathrm{DEC}}$ are the same in the uplink and downlink, but it is straightforward to assign them different values.

\subsection{Backhaul}

The backhaul is used to transfer uplink/downlink data between the BS and the core network. The power consumption of the backhaul is commonly modeled as the sum of two parts \cite{Tombaz2011a}: one load-independent and one load-dependent. The first part was already included in $P_{\mathrm{FIX}}$, while the load-dependent part is proportional to the average sum rate. Looking jointly at the downlink and uplink, the load-dependent term $P_{\mathrm{BH}}$ can be computed as \cite{Tombaz2011a}
\begin{align} \label{P_BH}
P_{\mathrm{BH}} =  \sum\limits_{k=1}^K \left( \mathbb{E} \left\{ R_k^{(\rm{ul})} + R_k^{(\rm{dl})} \right\}  \right)   P_{\mathrm{BT}}  \quad {\text{Watt}}
\end{align}
where $P_{\mathrm{BT}}$ is the backhaul traffic power (in Watt per bit/s).

\subsection{Linear Processing}

The transmitted and received vectors of information symbols at the BS are generated by transmit precoding and processed by receive combining, respectively. This costs \cite{Boyd2008a}
\begin{align}\label{P_LP}
P_\mathrm{LP} = \BW \Big(1 - \frac{(\tau^{\rm{(ul)}} + \tau^{\rm{(dl)}} ) K}{\CU}\Big) \frac{2MK}{L_{\rm{BS}}}  + P_\mathrm{LP-C} \quad {\text{Watt}}
\end{align}
where the first term describes the power consumed by making one matrix-vector multiplication per data symbol. The second term, $P_\mathrm{LP-C}$, accounts for the power required for the computation of $\mathbf{G}$ and $\mathbf{V}$. The precoding and combining matrices are computed once per coherence block and the complexity depends strongly on the choice of processing scheme. Since $\mathbf{G} = \mathbf{V}$ is a natural choice (except when the uplink and downlink are designed very differently), we only need to compute one of them and thereby reduce the computational complexity. If MRT/MRC is used, we only need to normalize each column of $\vect{H}$. This requires approximately
\begin{align}
P_\mathrm{LP-C}^{(\mathrm{MRT/MRC})} = \frac{\BW}{\CU} \frac{3MK}{L_{\rm{BS}}} \quad {\text{Watt}}
\end{align}
which was calculated using the arithmetic operations for standard linear algebra operations in \cite{Boyd2008a}.
On the other hand, if ZF processing is selected, then approximately
\begin{align}\label{ZF_power}
P_\mathrm{LP-C}^{(\mathrm{ZF})} = \frac{\BW}{\CU} \left( \frac{K^3}{3 L_{\rm{BS}}} + \frac{3 MK^2 + MK}{L_{\rm{BS}}} \right) \quad {\text{Watt}}
\end{align}
is consumed, if the channel matrix inversion implementation is based on standard Cholesky factorization and back-substitution \cite{Boyd2008a}. The computation of optimal MMSE processing is more complicated since the power allocation in \eqref{MRC.1} is a fixed-point equation that needs to be iterated until convergence. Such fixed-point iterations usually converge very quickly, but for simplicity we fix the number of iterations to some predefined number $Q$. This requires $P_\mathrm{LP-C}^{(\mathrm{MMSE})} = Q \, P_\mathrm{LP-C}^{(\mathrm{ZF})}$ Watt since the operations in each iteration are approximately the same as in ZF.

\section{Energy Efficiency Optimization with ZF Processing}\label{optimization_ZF}

The EE optimization in Problem \ref{problem-statement} is solved in this section under the assumption that ZF processing is employed in the uplink and downlink. This choice is not only motivated by analytic convenience but also because the numerical results (provided later) show that it is close-to-optimal. A similar analysis for MRC was conducted in \cite{Mukherjee2014a}, after the submission of this paper.

For ZF processing, Problem \ref{problem-statement} reduces to
\begin{equation} \label{problem-statement-ZF}
\maximize{\substack{ M \in \mathbb{Z}_+, \, K \in \mathbb{Z}_+, \,  \PP \geq 0 \\ M \ge K+1} } \quad \mathrm{EE}^{(\rm{ZF})} =  \frac{  K \Big( 1-\frac{\tau_{\mathrm{sum}} K}{\CU} \Big) \bar{R} }{ \frac{\BW \sigma^2 \PP \mathcal{S}_\mathbf{x}}{\eta} K + P_{\mathrm{CP}}^{\rm{(ZF)}}}
\end{equation}
where we have introduced the notation
\begin{equation}
\tau_{\mathrm{sum}} = \tau^{\rm{(ul)}} + \tau^{\rm{(dl)}},
\end{equation}
used the expression in \eqref{power_DL.10}, and the fact that 
 \begin{equation}
 \mathbb{E}\{ R_k^{(\rm{dl})} \} + \mathbb{E}\{ R_k^{(\rm{ul})} \} = R_k^{(\rm{dl})} + R_k^{(\rm{ul})} = ( 1-\frac{\tau_{\mathrm{sum}} K}{\CU} ) \bar{R}
 \end{equation}
and
 \begin{equation}\label{P_CP_ZF}
P_{\mathrm{CP}}^{\rm{(ZF)}} = P_{\mathrm{FIX}} + P_\mathrm{TC} + P_\mathrm{CE} + P_\mathrm{C/D} + P_\mathrm{BH} + P_\mathrm{LP}^{\rm{(ZF)}}
 \end{equation}
 with $P_\mathrm{LP}^{\rm{(ZF)}}$ being given by \eqref{P_LP} after replacing $P_\mathrm{LP-C}$ with $P_\mathrm{LP-C}^{\rm{(ZF)}}$ from \eqref{ZF_power}.

   \begin{table}[t]
\renewcommand{\arraystretch}{1.}
\caption{Circuit power coefficients for ZF processing}
\label{table_coefficients} 
\centering
\begin{tabular}{|l||l|}
\hline
\bfseries \quad Coefficients $\{\mathcal C_i\}$& \bfseries \quad Coefficients $\mathcal{A}$ and $\{\mathcal D_i\}$ \\
\hline

$\mathcal C_0 =P_{\mathrm{FIX}}  + P_{\mathrm{SYN}}$ &  $\mathcal{A} = P_{\mathrm{COD}} + P_{\mathrm{DEC}} + P_{\mathrm{BT}}$ \\

$\mathcal C_1=P_{\mathrm{UE}}$ & $\mathcal D_0=P_{\mathrm{BS}}$ \\

$\mathcal C_2=\frac{4 B \tau^{\rm{(dl)}}}{U L_{\rm{UE}}}$ & $\mathcal D_1=\frac{B}{L_{\rm{BS}}}(2+\frac{1}{\CU})$ \\

$\mathcal C_3=\frac{B}{3UL_{\rm{BS}}}$ &  $\mathcal D_2= \frac{B}{U L_{\rm{BS}}}(3-2\tau^{\rm{(dl)}})$ \\

\hline
\end{tabular}\vspace{-0.3cm}
\end{table}

 For notational convenience, we introduce the constant coefficients $\mathcal{A}$, $\{\mathcal{C}_i\}$, and $\{\mathcal{D}_i\}$ reported in Table \ref{table_coefficients}. These coefficients collect all the different terms in \eqref{P_{TC}}, \eqref{P_CE}, \eqref{P_BH}, \eqref{P_C/D}, and \eqref{P_LP} and allow us to rewrite $P_{\mathrm{CP}}^{\rm{(ZF)}}$ in \eqref{P_CP_ZF} in the more compact form
\begin{equation} \label{P_CP_compact-form}
P_{\mathrm{CP}}^{\rm{(ZF)}} = \sum\limits_{i=0}^3 \mathcal{C}_i K^i + M \sum\limits_{i=0}^2 \mathcal{D}_i K^i + \mathcal{A} K \Big( 1-\frac{\tau_{\mathrm{sum}} K}{\CU} \Big) \bar{R}
\end{equation}
where we recall that $\bar R$ is given by \eqref{constant_rate} and, thus, is also a function of $(M,K,\PP)$. Plugging \eqref{P_CP_compact-form} into \eqref{problem-statement-ZF} yields\footnote{Observe that the subsequent analysis is generic with respect to the coefficients $\mathcal{A}$, $\{\mathcal{C}_i\}$, and $\{\mathcal{D}_i\}$, while we use the hardware characterization in Table \ref{table_coefficients} for simulations in Section \ref{numerical_results}.}
\begin{align} \label{EE_ZF} 
&\mathrm{EE}^{(\rm{ZF})} = \\ & \frac{  K \Big( 1-\frac{\tau_{\mathrm{sum}} K}{\CU} \Big) \bar{R} }{ \frac{\BW \sigma^2 \PP \mathcal{S}_\mathbf{x}}{\eta} K + \sum\limits_{i=0}^3 \mathcal{C}_i K^i + M \sum\limits_{i=0}^2 \mathcal{D}_i K^i + \mathcal{A} K \Big( 1-\frac{\tau_{\mathrm{sum}}  K}{\CU} \Big) \bar{R} }.\nonumber
\end{align}
In the following, we aim at solving \eqref{problem-statement-ZF} for fixed $\mathcal{A}$, $\{\mathcal{C}_i\}$, and $\{\mathcal{D}_i\}$. In doing so, we first derive a closed-form expression for the EE-optimal value of either $M$, $K$, or $\PP$, when the other two are fixed.
This does not only bring indispensable insights on the interplay between these parameters and the coefficients $\mathcal{A}$, $\{\mathcal{C}_i\}$, and $\{\mathcal{D}_i\}$, but provides the means to solve the problem by an alternating optimization algorithm. All the mathematical proofs are given in the appendix.

\subsection{Preliminary Definition and Results}
\label{sec:preliminaries}

\begin{definition}  The Lambert W function is denoted by $W(x)$ and defined by the equation $x = W(x) e^{W(x)}$ for any $x \in \mathbb{C}$. \label{def:Lambert}
\end{definition}

\begin{lemma} \label{lemma:optimization-EE}
Consider the optimization problem
\begin{equation} \label{eq:lemma-opt-problem}
\maximize{z > -\frac{a}{b} } \quad \frac{g \log(a+bz) }{c + d z + h \log(a+bz)}
\end{equation}
with constant coefficients $a \in \mathbb{R}$, $c,h \geq 0$, and $b,d, g>0$. The unique solution to \eqref{eq:lemma-opt-problem} is
\begin{equation} \label{eq:lemma-opt-problem-solution}
z^{\star} = \frac{e^{W \left(\frac{bc}{de} -\frac{a}{e}\right)+1}-a}{b}.
\end{equation}
\end{lemma}

\begin{lemma} \label{lemma:lambert-bounds}
The Lambert W function $W(x)$ is an increasing function for $x \geq 0$ and satisfies the inequalities
\begin{equation}
e\frac{x}{\ln (x)} \leq  e^{W(x)+1} \leq (1+e)\frac{x}{\ln(x)} \quad \quad \text{for} \;\; x \geq e.
\end{equation}
\end{lemma}

The above lemma easily follows from the results and inequalities in \cite{Hoorfar2008a} and implies that $e^{W(x)+1}$ is approximately equal to $e$ for small $x$ (i.e., when $\ln (x)\approx x$) whereas it increases almost linearly with $x$ when $x$ takes large values. In other words,
\begin{align} \label{eq:approx1}
e^{W(x)+1} &\approx e \quad \text{for small values of $x$},\\
e^{W(x)+1} &\approx x \quad \text{for large values of $x$}. \label{eq:approx2}
\end{align}

Lemma \ref{lemma:optimization-EE} is used in this section to optimize the EE, while \eqref{eq:approx1} and \eqref{eq:approx2} are useful in the subsequent discussions to bring insights on how solutions in the form of $z^{\star}$ in \eqref{eq:lemma-opt-problem-solution} behave.

\subsection{Optimal Number of Users}

We start by looking for the EE-optimal value of $K$ when $M$ and $\PP$ are given. For analytic tractability, we assume that the sum SINR $\PP K$ (and thereby the PA power) and the number of BS antennas per UE, $\frac{M}{K}$, are kept constant and equal to $\PP K = \bar \PP$ and $\frac{M}{K} = \bar \beta$
with $\bar \PP > 0$ and $\bar \beta > 1$. The gross rate is thus fixed at $ \bar c = B \log ( 1 + { \bar \PP(\bar \beta-1)})$.
We have the following result.

\begin{theorem} \label{theorem-optimal-K}
Suppose $\mathcal{A}$, $\{\mathcal{C}_i\}$, and $\{\mathcal{D}_i\}$ are non-negative and constant.
For given values of $\bar \PP $ and $\bar \beta$, the number of UEs that maximize the EE metric is
\begin{equation}  \label{K_star}
K^\star = \max_{\ell} \left\lfloor K_\ell^{(o)}\right\rceil
\end{equation}
where the quantities $\{K_\ell^{(o)}\}$ denote the real positive roots of the quartic equation
\begin{equation} \label{eq:polynomial-K}
K^4 - \frac{2 \CU}{\tau_{\mathrm{sum}} } K^3 - \mu_1 K^2 - 2\mu_0K +  \frac{\CU \mu_0}{\tau_{\mathrm{sum}} }=0
\end{equation}
where $\mu_1 = \frac{\frac{\CU}{\tau_{\mathrm{sum}} }(\mathcal C_2 + \bar \beta \mathcal D_1) + \mathcal C_1 + \bar \beta \mathcal D_0 }{\mathcal C_{3} + \bar \beta \mathcal D_{2}}$ and $\mu_0= \frac{\mathcal C_0 + \frac{ \BW \sigma^2 \mathcal{S}_{\mathbf{x}}}{\eta}\bar \PP}{\mathcal C_{3} + \bar \beta \mathcal D_{2}}$.
\end{theorem}

This theorem shows that the optimal $K$ is a root to the quartic polynomial given in \eqref{eq:polynomial-K}. The notation $\lfloor \cdot \rceil$ in \eqref{K_star} says that the optimal value $K^\star$ is either the closest smaller or closest larger integer to $K_\ell^{(o)}$, which is easily determined by comparing the corresponding EE. A basic property in linear algebra is that quartic polynomials have exactly 4 roots (some can be complex-valued) and there are generic closed-form root expressions \cite{Shmakov2011a}. However, these expressions are very lengthy and not given here for brevity---in fact, the closed-form expressions are seldom used because there are simple algorithms to find the roots with higher numerical accuracy  \cite{Shmakov2011a}.

To gain insights on how $K^\star$ is affected by the different parameters, assume that the power consumption required for linear processing and channel estimation are both negligible (i.e., $P_\mathrm{CE}=P_{\mathrm{LP}}^{\rm{(ZF)}} \approx 0$). This case is particularly relevant as $P_\mathrm{CE}$ and $P_{\mathrm{LP}}^{\rm{(ZF)}}$ essentially decrease with the computational efficiencies $L_{\rm{BS}}$ and $L_{\rm{UE}}$, which are expected to increase rapidly in the future. Then, the following result is of interest.

\begin{corollary} \label{cor:optimal-K}
If $P_\mathrm{CE}$ and $P_{\mathrm{LP}}^{\rm{(ZF)}}$ are both negligible, then $
K^\star$ in \eqref{K_star} can be approximated as
\begin{equation} \label{eq-optimal-K-special-case}
K^\star \approx \left\lfloor\mu\left(\sqrt{ 1 + \frac{\CU}{\tau_{\mathrm{sum}}  \mu}    } - 1\right)\right\rceil
\end{equation}
with
\begin{equation}\label{eq-mu}
\mu = \frac{\mathcal{C}_0 + \frac{ \BW \sigma^2 \mathcal{S}_{\mathbf{x}}}{\eta}\bar \PP}{\mathcal{C}_1 + \bar \beta \mathcal{B}_0} =
\frac{P_{\mathrm{FIX}}  + P_{\mathrm{SYN}} + \frac{ \BW \sigma^2 \mathcal{S}_{\mathbf{x}}}{\eta}\bar \PP}{ P_{\mathrm{UE}} + \bar \beta P_{\mathrm{BS}}}.
\end{equation}
\end{corollary}

From \eqref{eq-optimal-K-special-case} and \eqref{eq-mu}, it is seen that  $K^\star$ is a decreasing function of the terms $\{P_{\mathrm{UE}},  P_{\mathrm{BS}}\}$ that are increasing with $K$ and/or $M$ in \eqref{P_CP}. On the contrary, $K^\star$ is an increasing function of the terms in \eqref{P_CP} that are independent of $K$ and $M$. This amounts to saying that the number of UEs increases with $\{P_{\mathrm{FIX}}, P_{\mathrm{SYN}}\}$ and $\mathcal{S}_{\mathbf{x}}$, as well as with the PA power (proportional to $\PP$) and the noise power $\sigma^2$. Looking at Example \ref{example:simulation-scenario}, $\mathcal{S}_{\mathbf{x}}$ increases proportionally to $d_{\max}^{\kappa}$ which means that a larger number of UEs must be served as the cell radius $d_{\max}$ increases.
Moreover, $K^\star$ is unaffected by the terms $\{P_{\mathrm{COD}}, P_{\mathrm{DEC}},P_{\mathrm{BT}}\}$, which are the ones that are multiplied with the average sum rate. The above results are summarized in the following corollaries.

\begin{corollary}
If the power consumptions for
linear processing and channel estimation are both negligible, then the optimal $K^\star$ decreases with the power per UE and BS antenna $\{P_{\mathrm{UE}},  P_{\mathrm{BS}}\}$, is unaffected by  the rate-dependent power $\{P_{\mathrm{COD}}, P_{\mathrm{DEC}}, P_{\mathrm{BT}}\}$, and increases with the fixed power $\{P_{\mathrm{FIX}}, P_{\mathrm{SYN}}\}$.
\end{corollary}

\begin{corollary}
A larger number of UEs must be served when the coverage area increases.
\end{corollary}

\subsection{Optimal Number of BS Antennas}

We now look for the $M \ge K+1$ that maximizes the EE in \eqref{EE_ZF} and have the following result.

\begin{theorem} \label{theorem-optimal-M}
For given values of $K$ and $\PP$, the number of BS antennas maximizing the EE metric can be computed as $M^\star =\left\lfloor M^{(o)} \right\rceil$ with
\begin{equation}  \label{eq:optimal_antennas}
M^{(o)}  = \frac{e^{W \left(\frac{\PP\left(\frac{  \BW \sigma^2 \mathcal{S}_{\mathbf{x}} }{\eta}\PP + \mathcal C^\prime\right)}{\mathcal D^\prime e} + \frac{\PP  K - 1 }{e } \right)+1} +\PP K - 1}{\PP}
\end{equation}
where $\mathcal{C}^\prime >0$ and $\mathcal{D}^\prime >0$ are defined as
\begin{equation}\label{B_prime}
\mathcal C^\prime = \frac{\sum_{i=0}^{3}  \mathcal C_{i} K^{i} }{K} \quad \text{and} \quad  \mathcal D^\prime = \frac{\sum_{i=0}^{2} \mathcal D_{i} K^{i}}{K }.
\end{equation}
\end{theorem}

Theorem \ref{theorem-optimal-M} provides explicit guidelines on how to select $M$ in a multi-user MIMO system to maximize EE. In particular, it provides the following fundamental insights.

\begin{corollary} \label{cor:optimal-M}
The optimal $M^\star$ does not depend on the rate-dependent power $\{P_{\mathrm{COD}},P_{\mathrm{DEC}}, P_{\mathrm{BT}}\}$ whereas it decreases with the power per BS antenna $\{P_{\mathrm{BS}}\}$ and  increases with the fixed power and UE-dependent power $\{P_{\mathrm{FIX}},P_{\mathrm{SYN}}, P_{\mathrm{UE}}\}$.
\end{corollary}

\begin{corollary} \label{cor:optimal-M-scaling}
The optimal $M^\star$ is lower bounded as
\begin{equation}  \label{eq:antenna-scaling}
M^\star \geq K+
\frac{ \frac{  \BW \sigma^2 \mathcal{S}_{\mathbf{x}} }{\eta \mathcal D^\prime }\PP + \frac{\mathcal C^\prime}{ \mathcal D^\prime}  + K -\frac{1 }{\PP } }{\ln (\PP) +   \ln \left(\frac{  \BW \sigma^2 \mathcal{S}_{\mathbf{x}} }{\eta \mathcal D^\prime }\PP + \frac{\mathcal C^\prime}{ \mathcal D^\prime}   + K -\frac{1 }{\PP }
\right) -1 }
  - \frac{1}{\PP}
\end{equation}
for moderately large values of $\PP$ (a condition is given in the proof). When $\PP$ grows large, we have
\begin{equation}
M^\star \approx \frac{  \BW \sigma^2 \mathcal{S}_{\mathbf{x}} }{2\eta \mathcal D^\prime } \frac{\PP}{\ln(\PP)}
\end{equation}
which is an almost linear scaling law.
\end{corollary}

\begin{corollary}
A larger number of antennas is needed as the size of the coverage area increases.
\end{corollary}
The above corollary follows from the observation that $M^\star$ increases almost linearly with $\mathcal{S}_{\mathbf{x}}$, which is a parameter that increases with the cell radius $d_{\max}^{\kappa}$ (as illustrated in Example \ref{example:simulation-scenario}).

\subsection{Optimal Transmit Power}

Recall that $\PP$ is proportional to the SINR, which is directly proportional to the PA/transmit power under ZF processing. Finding the EE-optimal total PA power amounts to looking for the value of $\PP$ in \eqref{power_DL.10} that maximizes \eqref{EE_ZF}. The solution is given by the following theorem.

\begin{theorem} \label{theorem-optimal-rho}
For given values of $M$ and $K$, the EE-optimal $\PP \geq 0$ can be computed as
\begin{equation} \label{eq:power-scaling}
\PP^\star = \frac{e^{W  \left( \frac{\eta}{ \BW \sigma^2 \mathcal{S}_{\mathbf{x}}}\frac{(M-K) (\mathcal{C}^\prime + M \mathcal{D}^\prime)}{e} -\frac{1}{e}\right)+1}-1}{M-K}
\end{equation}
with $\mathcal{C}^\prime >0 $ and $\mathcal{D}^\prime >0 $ given by \eqref{B_prime}.
\end{theorem}

Using Lemma \ref{lemma:lambert-bounds}, it turns out that the optimal $\PP^\star$ increases with $\mathcal{C}^\prime$ and $\mathcal{D}^\prime$, which were defined in \eqref{B_prime}, and thus with the coefficients in the circuit power model. Since the EE-maximizing total PA power with ZF processing is
$P_\mathrm{TX}^{(\rm{ZF})}  = \frac{\BW \sigma^2  \mathcal{S}_\mathbf{x}}{\eta} K \PP^\star$, the following result is found.

\begin{corollary}\label{cor:increase-power-with-circuit_powers}
The optimal transmit power does not depend on the rate-dependent power  $\{P_{\mathrm{COD}},P_{\mathrm{DEC}}, P_{\mathrm{BT}}\}$ whereas it increases with  the fixed power and the power per UE and BS antenna $\{P_{\mathrm{BS}},P_{\mathrm{FIX}},P_{\mathrm{SYN}},P_{\mathrm{UE}}\}$.
\end{corollary}

The fact that the optimal PA/transmit power \emph{increases} with $\{P_{\mathrm{BS}},P_{\mathrm{FIX}},P_{\mathrm{SYN}},P_{\mathrm{UE}}\}$ might seem a bit counterintuitive at first, but it actually makes much sense and can be explained as follows. If the fixed circuit powers are large, then higher PA power $P_{\mathrm{TX}}^{(\rm{ZF})}$ (and thus higher average rates) can be afforded in the system since  $P_{\mathrm{TX}}^{(\rm{ZF})}$ has small impact on the total power consumption.

It has recently been shown in \cite{Bjornson2014a,Hoydis2013a}, and \cite{Ngo2013a} that TDD systems permit a power reduction proportional to $1/M$ (or $1/\sqrt{M}$ with imperfect CSI) while maintaining non-zero rates as $M \rightarrow \infty$. Despite being a remarkable result and a key motivation for massive MIMO systems, Theorem \ref{theorem-optimal-rho} proves that this is \emph{not} the most energy-efficient strategy. In fact, the EE metric is  maximized by the opposite strategy of actually increasing the power with $M$.

\begin{corollary} \label{cor:increase-power-with-M}
The optimal $\PP^\star$ is lower bounded as
\begin{align}\label{power_opt_low_bound}
\PP^\star &\ge \frac{\frac{\eta (\mathcal{C}^\prime  + M\mathcal{D}^\prime )}{\BW \sigma^2 \mathcal{S}_{\mathbf{x}} }  - \frac{ \ln\left( \frac{\eta (M-K) (\mathcal{C}^\prime  + M\mathcal{D}^\prime )}{\BW  \sigma^2 \mathcal{S}_{\mathbf{x}} } -1 \right)}{\left(M-K\right)}  }{ \ln\left( \frac{\eta (M-K) (\mathcal{C}^\prime  + M\mathcal{D}^\prime )}{\BW  \sigma^2 \mathcal{S}_{\mathbf{x}} } -1 \right) -1}
\end{align}
for moderate and large values of $M$ (a condition is given in the proof) whereas
\begin{align}
\PP^\star \approx  \frac{ \eta \mathcal{D}^\prime  }{ 2  \BW \sigma^2 \mathcal{S}_{\mathbf{x}} } \frac{M}{\ln\left( M \right)}
\end{align}
when $M$ grows large.
\end{corollary}

The above corollary states that the total PA power $P_{\mathrm{TX}}^{(\rm{ZF})}$ required to maximize the EE metric increases approximately as ${M}/{\ln(M)}$, which is an almost linear scaling. The explanation is the same as for Corollary \ref{cor:increase-power-with-circuit_powers}: the circuit power consumption grows with $M$, thus we can afford using more transmit power to improve the rates before it becomes the limiting factor for the EE. Although the total transmit power increases with $M$, the average transmit power emitted per BS antenna (and per UE if we let $K$ scale linearly with $M$) actually decays as ${1}/{\ln(M)}$. Hence, the RF amplifiers can be gradually simplified with $M$.
The EE-maximizing per-antenna transmit power reduction is, nevertheless, much slower than the linear to quadratic scaling laws observed in \cite{Hoydis2013a} and \cite{Ngo2013a}, for the unrealistic case of no circuit power consumption.

\subsection{Joint and Alternating Optimization of $K$, $M$, and $\PP$.}
\label{subsec:joint-sequential}

Theorems \ref{theorem-optimal-K}--\ref{theorem-optimal-rho} provide simple closed-form expressions that enable EE-maximization by optimizing $K$, $M$, or $\PP$ separately when the other two parameters are fixed. However, the ultimate goal for a system designer is to find the joint global optimum. Since $K$ and $M$ are integers, the global optimum can be obtained by an exhaustive search over all reasonable combinations of the pair $(K,M)$ and computing the optimal power allocation for each pair using Theorem \ref{theorem-optimal-rho}. Since Theorem \ref{theorem-optimal-K} shows that the EE metric is quasi-concave when $K$ and $M$ are increased jointly (with a fixed ratio), one can increase $K$ and $M$ step-by-step and stop when the EE starts to decrease. Hence, there is no need to consider all integers.

Although feasible and utilized for simulations in Section \ref{numerical_results}, the brute-force joint optimization is of practical interest only for off-line cell planning, while a low-complexity approach is required to eventually take into account changes in the system settings (e.g., the user distribution or the path-loss model as specified by $\mathcal S_{\mathbf{x}}$). A practical solution in this direction is to optimize the system parameters sequentially according to a standard alternating optimization algorithm:

\begin{enumerate}
\item Assume that an initial set $(K,M,\PP)$ is given;
\item Update the number of UEs $K$ (and implicitly $M$ and $\PP$) according to Theorem \ref{theorem-optimal-K};
\item Replace $M$ with the optimal value from Theorem \ref{theorem-optimal-M};
\item Optimize the PA power through $\PP$ by using Theorem \ref{theorem-optimal-rho};
\item Repeat 2) -- 5) until convergence is achieved.
\end{enumerate}

Observe that the EE metric has a finite upper bound (for $\mathcal C_i>0$ and $\mathcal D_{i}>0$). Therefore, the alternating algorithm illustrated above monotonically converges to a local optimum for any initial set $(K,M,\PP)$, because the alternating updates of $K$, $M$, and $\PP$ may either increase or maintain (but not decrease) the objective function.
Convergence is declared when the integers $M$ and $K$ are left unchanged in an iteration.

\section{Extensions to Imperfect CSI and Multi-Cell Scenarios}\label{section:multicell}

The EE-optimal parameter values were derived in the previous section for a single-cell scenario with perfect CSI. In this section, we investigate to what extent the analysis can be extended to single-cell scenarios with imperfect CSI. We also derive a new achievable rate for symmetric multi-cell scenarios with ZF processing.

The following lemma gives achievable user rates in single-cell scenarios with imperfect CSI.

\begin{lemma} \label{lemma:ZF-rate-imperfect}
If approximate ZF detection/precoding is applied under imperfect CSI (acquired from pilot signaling and MMSE channel estimation), the average gross rate
\begin{equation} \label{eq:gross-rate-imperfect}
\bar{R} = \BW \log \left( 1+ \frac{\PP(M-K)}{ 1+\frac{1}{\tau^{(\rm{ul})}}+\frac{1}{\PP K \tau^{(\rm{ul})}} }  \right)
\end{equation}
is achievable using the same average PA power $\frac{\BW \sigma^2 \PP \mathcal{S}_\mathbf{x}}{\eta} K$ as in \eqref{power_DL.10}, where $\PP\geq 0$ is a parameter.
\end{lemma}
\begin{IEEEproof}
The proof is given in the appendix.
\end{IEEEproof}

The rate expression in \eqref{eq:gross-rate-imperfect} is different from \eqref{constant_rate} due to the imperfect CSI which causes unavoidable interference between the UEs. In particular, the design parameters $K$ and $\PP$ appear in both the numerator and denominator of the SINRs, while these only appeared in the numerator in \eqref{constant_rate}. Consequently, we cannot find the EE-optimal $K$ and $\PP$ in closed form under imperfect CSI. The optimal number of BS antennas can, however, be derived similarly to Theorem \ref{theorem-optimal-M}:
\begin{align}  \label{eq:optimal_antennas_imperfect}
& M^\star = \Bigg\lfloor \left(1+\frac{1}{\tau^{(\rm{ul})}}+\frac{1}{\PP K \tau^{(\rm{ul})}} \right) \times \\  &\frac{ \, e^{W \left(\frac{\PP\left(\frac{  \BW \sigma^2 \mathcal{S}_{\mathbf{x}} }{\eta}\PP + \mathcal C^\prime\right)}{\mathcal D^\prime e (1+\frac{1}{\tau^{(\rm{ul})}}+\frac{1}{\PP K \tau^{(\rm{ul})}})} + \frac{\PP  K - 1 }{e (1+\frac{1}{\tau^{(\rm{ul})}}+\frac{1}{\PP K \tau^{(\rm{ul})}})} \right)+1} +\PP K - 1}{\PP} \Bigg\rceil. \nonumber
\end{align}
Despite the analytic difficulties, Section \ref{numerical_results} shows numerically that the single-cell behaviors that were proved in Section \ref{optimization_ZF} are applicable also under imperfect CSI.

The analytic framework and observations of this paper can also be applied in multi-cell scenarios. To illustrate this, we consider a completely symmetric scenario where the system parameters $M$, $K$, and $\bar{R}$ are the same in all cells and optimized jointly. The symmetry implies that the cell shapes, user distributions, and propagation conditions are the same in all cells.

We assume that there are $J$ cells in the system. Let $\vect{x}_{jk}$ denote the position of the $k$th UE in cell $j$ and call $l_j(\vect{x})$ the average channel attenuation between a certain position $\vect{x} \in \mathbb{R}^2$ and the $j$th BS. The symmetry implies that the average inverse attenuation to the serving BS, $\mathcal{S}_{\mathbf{x}} = \mathbb{E} \left\{{ (l_j(\mathbf{x}_{jk})})^{-1}\right\}$, is independent of the cell index $j$. Moreover, we define
\begin{equation}
\mathcal{I}_{j \ell} = \mathbb{E}_{\vect{x}_{\ell k}} \left\{ \frac{l_j(\vect{x}_{\ell k})}{l_{\ell}(\vect{x}_{\ell k})} \right\}
\end{equation}
as the average ratio between the channel attenuation to another BS and the serving BS. This parameter describes the average interference that leaks from a UE in cell $\ell$ to the BS in cell $k$ in the uplink, and in the inverse direction in the downlink. The symmetry implies $\mathcal{I}_{j \ell} = \mathcal{I}_{\ell j}$.

The necessity of reusing pilot resources across cells causes pilot contamination (PC) \cite{Marzetta2010a}. To investigate its impact on the EE, we consider different pilot reuse patterns by defining $\mathcal{Q}_j \subset \{1,2,\ldots, J\}$ as the set of cells (including cell $j$) that use the same pilot sequences as cell $j$. For symmetry reasons, we let the cardinality $|\mathcal{Q}_j|$ be the same for all $j$. We also note that the uplink pilot sequence length is $K \tau^{(\rm{ul})}$, where $\tau^{(\rm{ul})} \geq {J}/{|\mathcal{Q}_j|}$ to account for the pilot reuse factor.
The average relative power from PC is $\mathcal{I}_{\rm{PC}} = \sum_{\ell \in \mathcal{Q}_j \setminus \{j\} } \mathcal{I}_{j \ell}$, while $\mathcal{I}= \sum_{\ell=1}^{J} \mathcal{I}_{j \ell}$ is the relative interference from all cells and $\mathcal{I}_{\rm{PC}^2} = \sum_{\ell \in \mathcal{Q}_j \setminus \{j\}} \mathcal{I}_{j \ell}^2$ is defined for later use. Note that these parameters are also independent of $j$ for symmetry reasons.

\begin{lemma} \label{lemma:ZF-rate-multicell}
If ZF detection/precoding is applied by treating channel uncertainty as noise, the average total PA power $\frac{\BW \sigma^2 \PP \mathcal{S}_\mathbf{x}}{\eta} K$ in \eqref{power_DL.10} achieves the average gross rate
\begin{align} \nonumber
& \bar{R} = \!\BW \times \\ & \log \! \Bigg( 1+ \frac{1}{\mathcal{I}_{\rm{PC}} + (1+\mathcal{I}_{\rm{PC}}+\frac{1}{\PP K \tau^{(\rm{ul})}}) \frac{(1+K \PP \mathcal{I}  )}{\PP(M-K)} - \frac{K  (1 + \mathcal{I}_{\rm{PC}^2} )}{M-K} }  \Bigg)  \label{eq:gross-rate-multicell}
\end{align}
in each cell, where $\PP \geq 0$ is a design parameter.
\end{lemma}
\begin{IEEEproof}
The proof is given in the appendix.
\end{IEEEproof}

The rate expression in \eqref{eq:gross-rate-multicell} for symmetric multi-cell scenarios (with imperfect CSI) is even more complicated than the single-cell imperfect CSI case considered in Lemma \ref{lemma:ZF-rate-imperfect}. All the design parameters $M$, $K$, and $\PP$ appear in both the numerator and denominator of the SINRs, which generally makes it intractably to find closed-form expressions for the EE-optimal parameter values. Indeed, this is the reason why we devoted Section \ref{optimization_ZF} to an analytically tractable single-cell scenario.  Nevertheless, we show in the next section that symmetric multi-cell scenarios behave similarly to single-cell scenarios, by utilizing the rate expression in \eqref{eq:gross-rate-multicell} for simulations.

\begin{table*}[t]
\renewcommand{\arraystretch}{1.}
\caption{Simulation Parameters}
\label{table_parameters_hardware}
\begin{tabular}{|c|c||c|c|}
\hline
\bfseries Parameter & \bfseries Value & \bfseries $\!\!\!\!\!$ Parameter $\!\!\!\!\!$ & \bfseries Value\\
\hline

Cell radius (single-cell): $d_{\max}$ & $250$ m & Fraction of downlink transmission: $\zeta^{(\rm{dl})}$ & $0.6$ \\

Minimum distance: $d_{\min}$ & $35$ m &  Fraction of uplink transmission: $\zeta^{(\rm{ul})}$ & $0.4$ \\

Large-scale fading model: $l(\vect{x})$ & ${10^{-3.53}}/{\| \vect{x} \|^{3.76}}$ &  PA efficiency at the BSs: $\eta^{(\rm{dl})}$ & $0.39$ \\

Transmission bandwidth: $B$ & $20$ MHz & PA efficiency at the UEs: $\eta^{(\rm{ul})}$ & $0.3$ \\

Channel coherence bandwidth: $\BW_{\rm{C}}$ & $180$ kHz & Fixed power consumption (control signals, backhaul, etc.): $P_{\mathrm{FIX}}$ & $18$ W \\

Channel coherence time: $T_{\rm{C}}$ & $10$ ms & Power consumed by local oscillator at BSs: $P_{\mathrm{SYN}}$ & $2$ W \\

Coherence block (symbols): $\CU$  & $1800$ & Power required to run the circuit components at a BS: $P_{\mathrm{BS}}$ & $1$ W  \\

Total noise power: $B \sigma^2$ & $-96$ dBm & Power required to run the circuit components at a UE: $P_{\mathrm{UE}}$ & $0.1$ W \\

Relative pilot lengths: $\tau^{\rm{(ul)}},\tau^{\rm{(dl)}}$ & $1$ &  Power required for coding of data signals: $P_{\mathrm{COD}}$ & $\!\!\!\!$  $0.1$ W/(Gbit/s) $\!\!\!\!$\\

Computational efficiency at BSs: $L_{\rm{BS}}$ & $12.8$  Gflops/W & Power required for decoding of data signals: $P_{\mathrm{DEC}}$ & $\!\!\!\!$  $0.8$ W/(Gbit/s) $\!\!\!\!$\\

Computational efficiency at UEs: $L_{\rm{UE}}$ & $5$ Gflops/W &  Power required for backhaul traffic: $P_{\mathrm{BT}}$ &  $\!\!\!\!$ $0.25$ W/(Gbit/s) $\!\!\!\!$\\

\hline
\end{tabular}\vspace{-0.3cm}
\end{table*}

\section{Numerical Results}\label{numerical_results}

This section uses simulations to validate the system design guidelines obtained in Section \ref{optimization_ZF} under ZF processing and to make comparisons with other processing schemes. We provide numerical results under both perfect and imperfect CSI, and for both single-cell and multi-cell scenarios. Analytic results were used to simulate ZF, while Monte Carlo simulations with random user locations and small-scale fading were conducted to optimize EE with other schemes.

To compute the total power consumption in a realistic way, we use the hardware characterization described in Section \ref{circuit_power_model}. We first consider the single-cell simulation scenario in Example \ref{example:simulation-scenario} (i.e., a circular cell with radius 250 m) and assume operation in the 2 GHz band. The corresponding simulation parameters are given in Table \ref{table_parameters_hardware} and are inspired by a variety of prior works: the 3GPP propagation environment defined in \cite{LTE2010b}, RF and baseband power modeling from \cite{EARTH_D23,Kumar2011a,Mezghani2011a,Kang2011a}, backhaul power according to \cite{Tombaz2012a}, and the computational efficiencies are from \cite{Yang2013a,Parker2009a}. The simulations were performed using Matlab and the code is available for download at \url{https://github.com/emilbjornson/is-massive-MIMO-the-answer}, which enables reproducibility as well as simple testing of other parameter values.

\subsection{Single-Cell Scenario}

\begin{figure}
\begin{center}
\includegraphics[width=\columnwidth]{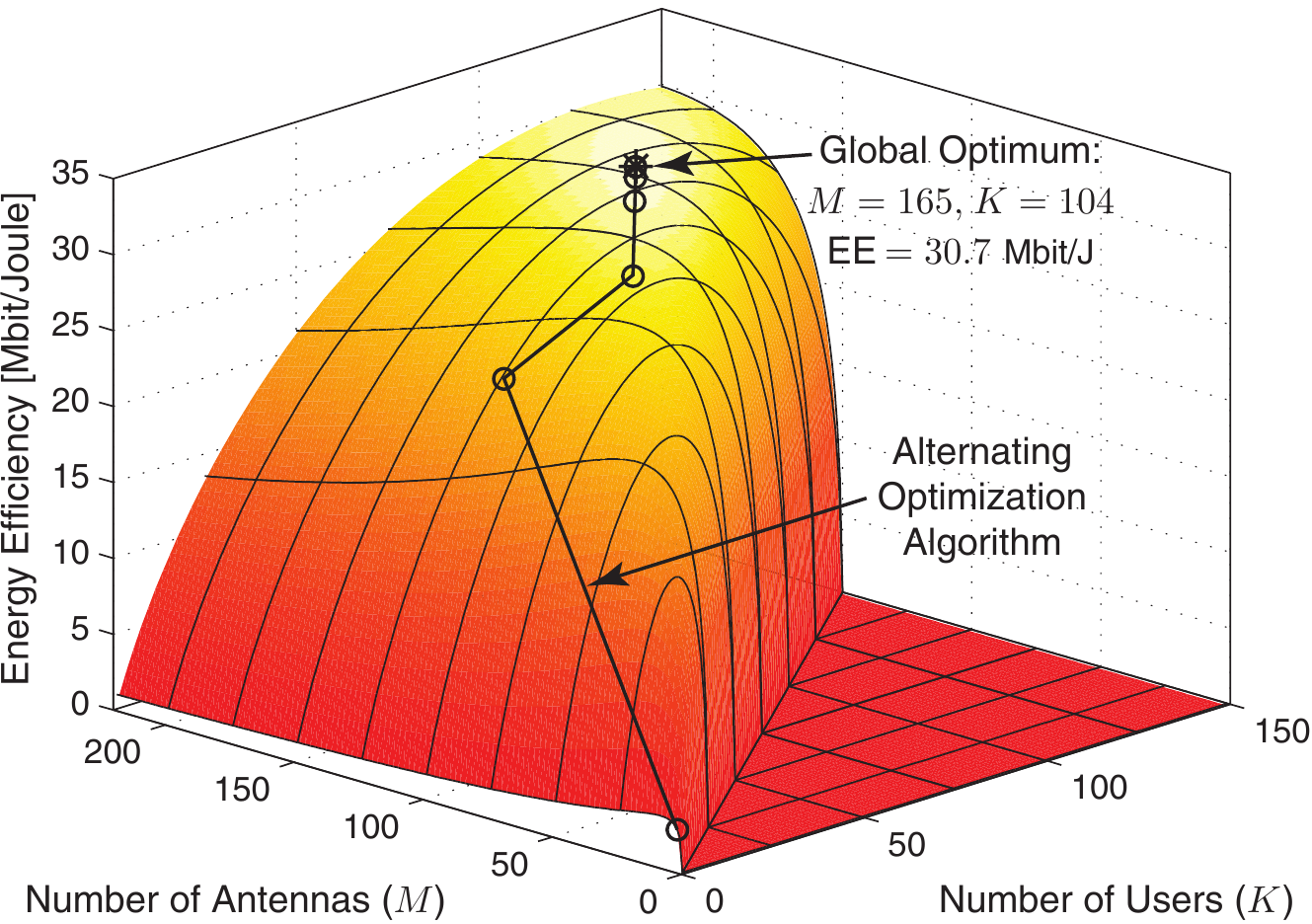}
\end{center}
\vspace{-0.3cm}
\caption{Energy efficiency (in Mbit/Joule) with ZF processing in the single-cell scenario. The global optimum is star-marked and the surroundings are white. The convergence of the proposed alternating optimization algorithm is indicated with circles.} \label{figure_3d_ZF}
\end{figure}

Fig.~\ref{figure_3d_ZF} shows the set of achievable EE values with perfect CSI, ZF processing, and for different values of $M$ and $K$ (note that $M \geq K+1$ in ZF). Each point uses the EE-maximizing value of $\PP$ from Theorem \ref{theorem-optimal-rho}. The figure shows that there is a global EE-optimum at $M = 165$ and $K = 104$, which is achieved by $\PP=0.8747$ and the practically reasonable spectral efficiency $5.7644$ bit/symbol (per UE). The optimum is clearly a massive MIMO setup, which is noteworthy since it is the output of an optimization problem where we did not restrict the system dimensions whatsoever. The surface in Fig.~\ref{figure_3d_ZF} is concave and quite smooth; thus, there is a variety of system parameters that provides close-to-optimal EE and the results appear to be robust to small changes in the circuit power coefficients. The alternating optimization algorithm from Section \ref{subsec:joint-sequential} was applied with a starting point in $(M,K,\PP) =(3,1,1)$. The iterative progression is shown in Fig.~\ref{figure_3d_ZF} and the algorithm converged after 7 iterations to the global optimum.

\begin{figure}
\begin{center}
\includegraphics[width=\columnwidth]{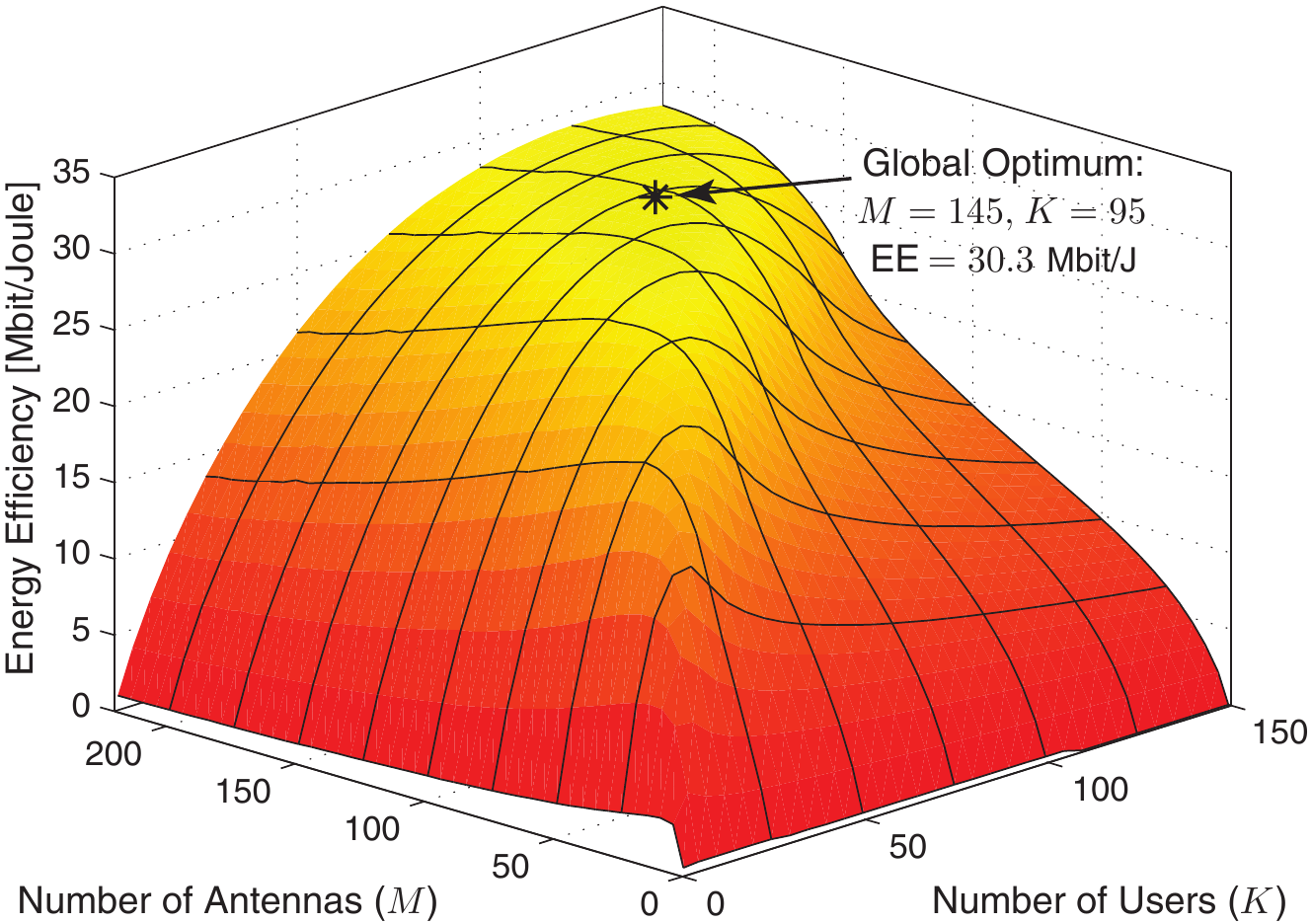}
\end{center}
\vspace{-0.3cm}
\caption{Energy efficiency (in Mbit/Joule) with MMSE processing  in the single-cell scenario.} \label{figure_3d_MMSE}
\end{figure}

\begin{figure}
\begin{center}
\includegraphics[width=\columnwidth]{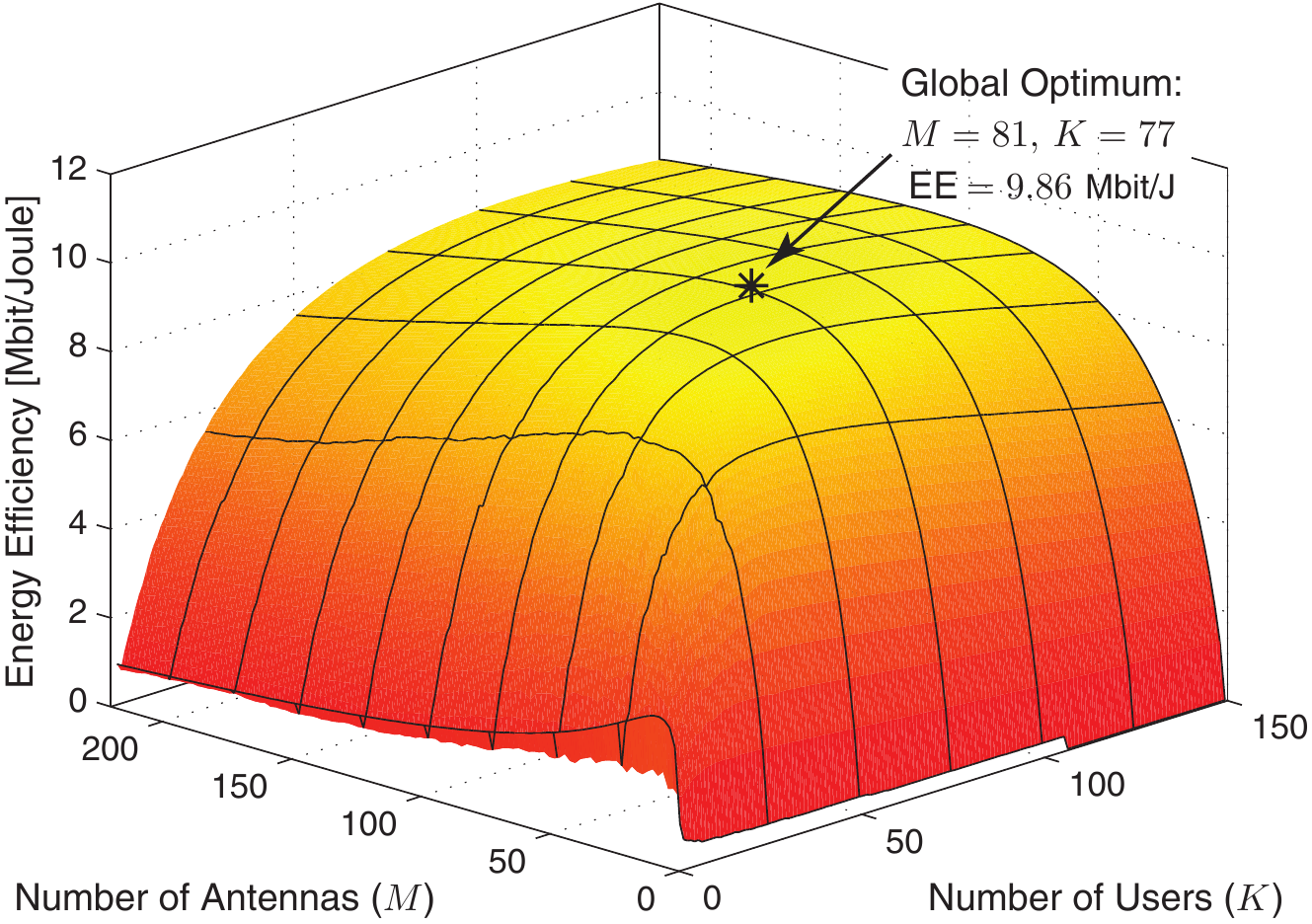}
\end{center}
\vspace{-0.3cm}
\caption{Energy efficiency (in Mbit/Joule) with MRT/MRC processing in the single-cell scenario.}\label{figure_3d_MRT}
\end{figure}

\begin{figure}
\begin{center}
\includegraphics[width=\columnwidth]{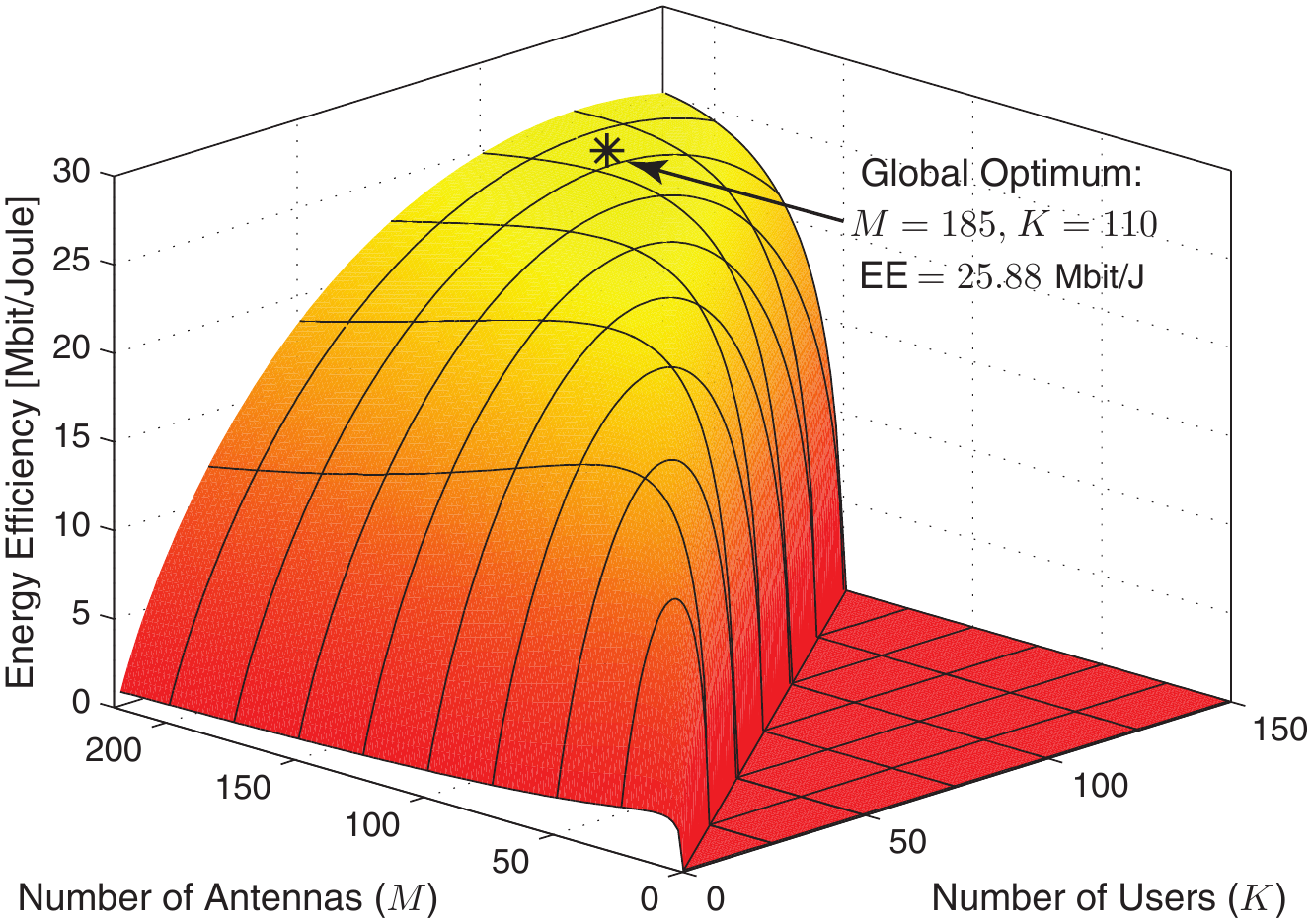}
\end{center}
\vspace{-0.3cm}
\caption{Energy efficiency (in Mbit/Joule) with ZF processing in the single-cell scenario with imperfect CSI.} \label{figure_3d_ZFimperfect}
\end{figure}

For comparisons, Fig.~\ref{figure_3d_MMSE} shows the corresponding set of achievable EE values under MMSE processing (with $Q=3$), Fig.~\ref{figure_3d_MRT} illustrates the results for MRT/MRC processing, and Fig.~\ref{figure_3d_ZFimperfect} considers ZF processing under imperfect CSI. The MMSE and MRT/MRC results were generated by Monte Carlo simulations, while the ZF results were computed using the expression in Lemma \ref{lemma:ZF-rate-imperfect}. Although MMSE processing is optimal from a throughput perspective, we observe that ZF processing achieves higher EE. This is due to the higher computational complexity of MMSE. The difference is otherwise quite small. MMSE has the (unnecessary) benefit of also handling $M<K$. ZF with imperfect CSI has a similar behavior as ZF and MMSE with perfect CSI, thus the analysis in Section \ref{optimization_ZF} has a bearing also on realistic single-cell systems.

Interestingly, MRT/MRC processing gives a very different behavior: the EE optimum is much smaller than with ZF/MMSE and is achieved at $M = 81$ and $K = 77$.\footnote{Single-user transmission was optimal for MRT in our previous work \cite{Bjornson2014b}, where we used another power consumption model. As compared to \cite{Bjornson2014b}, we have increased the backhaul power consumption (based on numbers from \cite{Tombaz2012a}) and made the coding/decoding power proportional to the rates instead of the number of UEs.} This can still be called a massive MIMO setup since there is a massive number of BS antennas, but it is a degenerative case where $M$ and $K$ are almost equal and thus the typical asymptotic massive MIMO properties from \cite{Marzetta2010a,Ngo2013a} will not hold. The reason for $M \approx K$ is that MRT/MRC operates under strong inter-user interference, thus the rate per UE is small and it makes sense to schedule as many UEs as possible (to crank up the sum rate). The signal processing complexity is lower than with ZF for the same $M$ and $K$, but the power savings are not big enough to compensate for the lower rates. To achieve the same rates as with ZF, MRT/MRC requires $M \gg K$ which would drastically increase the computational/circuit power and not improve the EE.

Looking at the respective EE-optimal operating points, we can use the formulas in Section \ref{circuit_power_model} to compute the total complexity of channel estimation, computing the precoding/combining matrices, and performing precoding and receive combining: it becomes 710 Gflops with ZF, 239 Gflops with MRT/MRC, and 664 Gflops with MMSE. These numbers are all within a realistic range and a vast majority of the computations can be parallelized for each antenna. Despite its larger number of BS antennas and UEs, ZF processing only requires $3 \times$ more operations than MRT/MRC. This is because the total complexity is dominated by performing precoding and receive combining on every vector of data symbols, while the computation of the precoding matrix (which scales as $\mathcal{O}(K^3+MK^2)$ for ZF) only occurs once per coherence block.

\begin{figure}
\begin{center}
\includegraphics[width=\columnwidth]{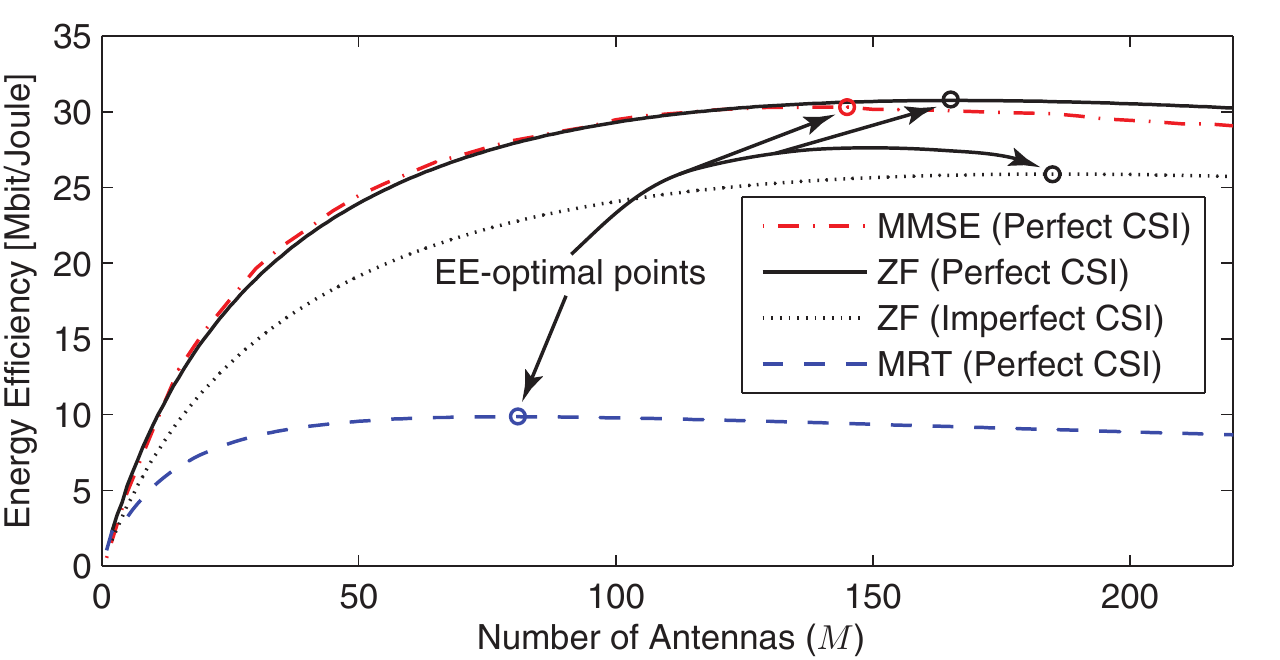}
\end{center}
\vspace{-0.3cm}
\caption{Maximal EE for different number of BS antennas and different processing schemes in the single-cell scenario.} \label{figure_EE}
\end{figure}

To further compare the different processing schemes, Fig.~\ref{figure_EE} shows the maximum EE as a function of the number of BS antennas. Clearly, the similarity between MMSE and ZF shows an optimality of operating at high SNRs (where these schemes are almost equal).

\begin{figure}
\begin{center}
\includegraphics[width=\columnwidth]{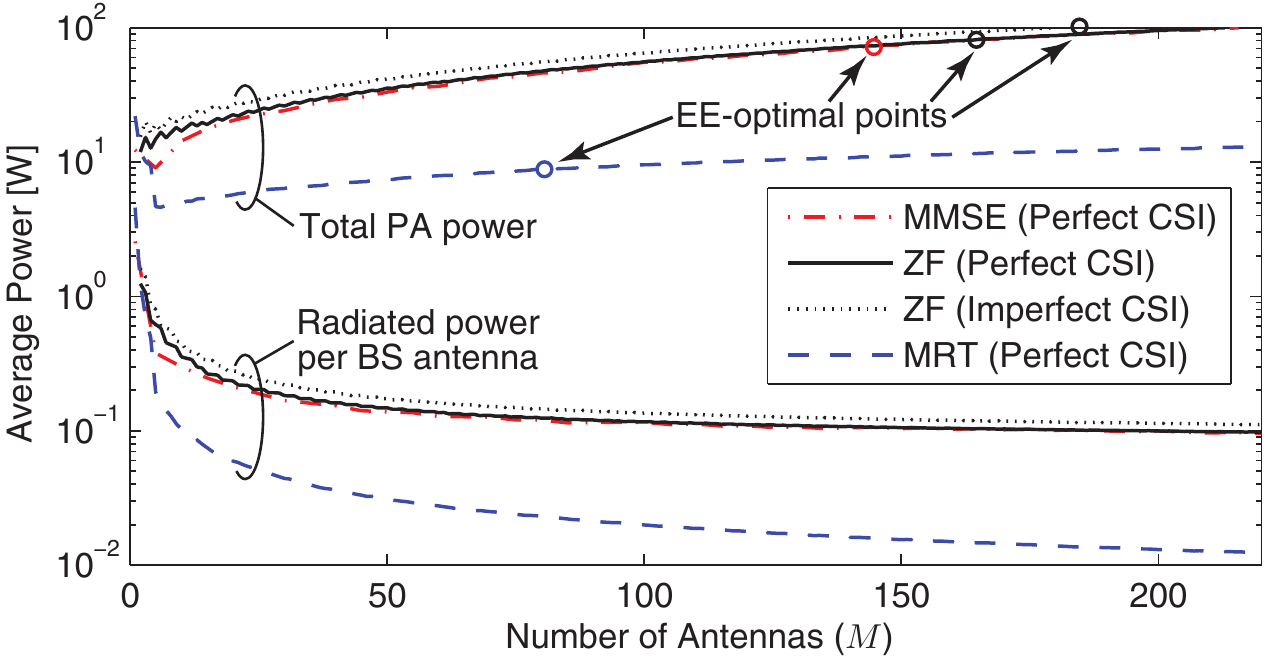}
\end{center}
\vspace{-0.3cm}
\caption{Total PA power at the EE-maximizing solution for different number of BS antennas in the single-cell scenario. The radiated power per BS antenna is also shown.} \label{figure_power}
\end{figure}

Next, Fig.~\ref{figure_power} shows the total PA power that maximizes the EE for different $M$ (using the corresponding optimal $K$). For all the considered processing schemes, the most energy-efficient strategy is to \emph{increase} the transmit power with $M$. This is in line with Corollary \ref{cor:increase-power-with-M} but stands in contrast to the results in \cite{Hoydis2013a} and \cite{Ngo2013a}, which indicated that the transmit power should be decreased with $M$. However, Fig.~\ref{figure_power} also shows that the transmit power \emph{per} BS antenna decreases with $M$. The downlink transmit power with ZF and MMSE precoding is around 100 mW/antenna, while it drops to 23 mW/antenna with MRT since it gives higher interference and thus makes the system interference-limited at lower power. These numbers are much smaller than for conventional macro BSs (which operate at around $40\cdot 10^3$ mW/antenna \cite{LTE2010b}) and reveals that the EE-optimal solution can be deployed with low-power UE-like RF amplifiers. Similar transmit power levels are observed for the UEs in the uplink, but are not included in Fig.~\ref{figure_power} for brevity.

\begin{figure}
\begin{center}
\includegraphics[width=\columnwidth]{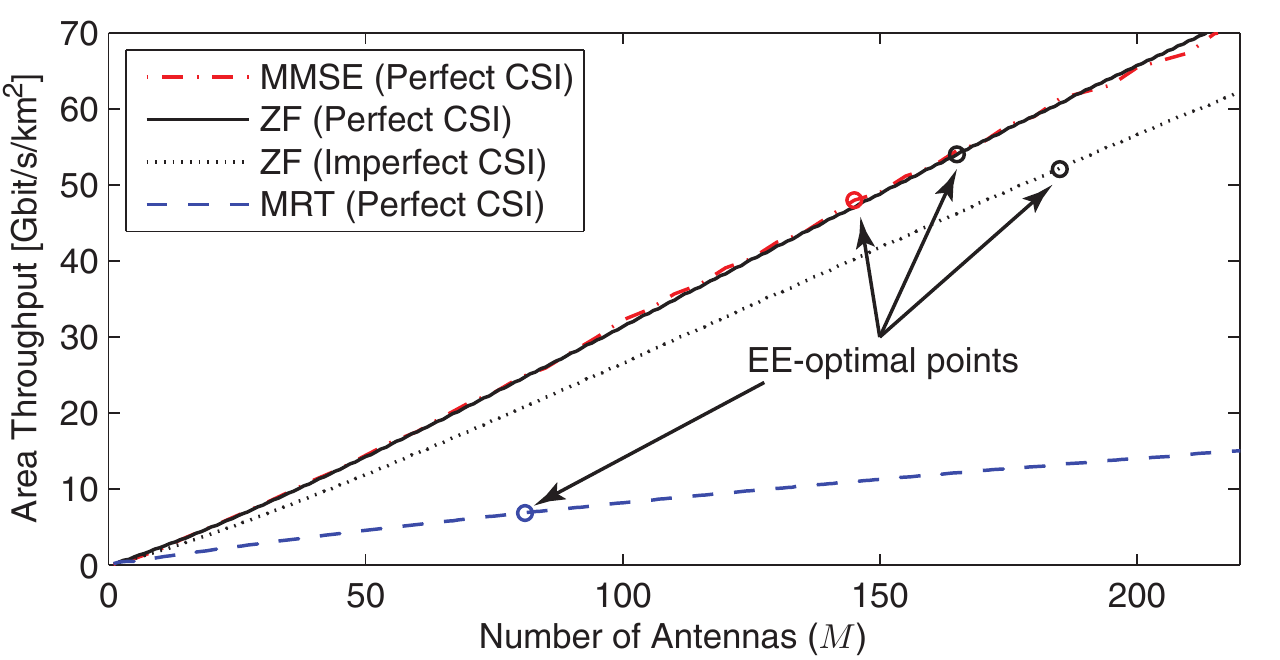}
\end{center}
\vspace{-0.3cm}
\caption{Area throughput at the EE-maximizing solution for different number of BS antennas in the single-cell scenario.} \label{figure_rates}
\end{figure}

Finally, Fig.~\ref{figure_rates} shows the area throughput (in $\mathrm{Gbit/s/km}^2$) that maximizes the EE for different $M$.
We consider the same processing schemes as in Figs.~\ref{figure_EE} and~\ref{figure_power}. Recall from Fig.~\ref{figure_EE} that there was a 3-fold improvement in optimal EE for ZF and MMSE processing as compared to MRT/MRC. Fig.~\ref{figure_rates} shows that there is simultaneously an 8-fold improvement in area throughput. The majority of this gain is achieved also under imperfect CSI, which shows that massive MIMO with proper interference-suppressing precoding can achieve both great energy efficiency and unprecedented area throughput. In contrast, it is wasteful to deploy a large number of BS antennas and then co-process them using a MRT/MRC processing scheme that is severely limiting both the energy efficiency and area throughput.

\subsection{Multi-Cell Scenario}

\begin{figure}[t!]
\begin{center}
\includegraphics[width=\columnwidth]{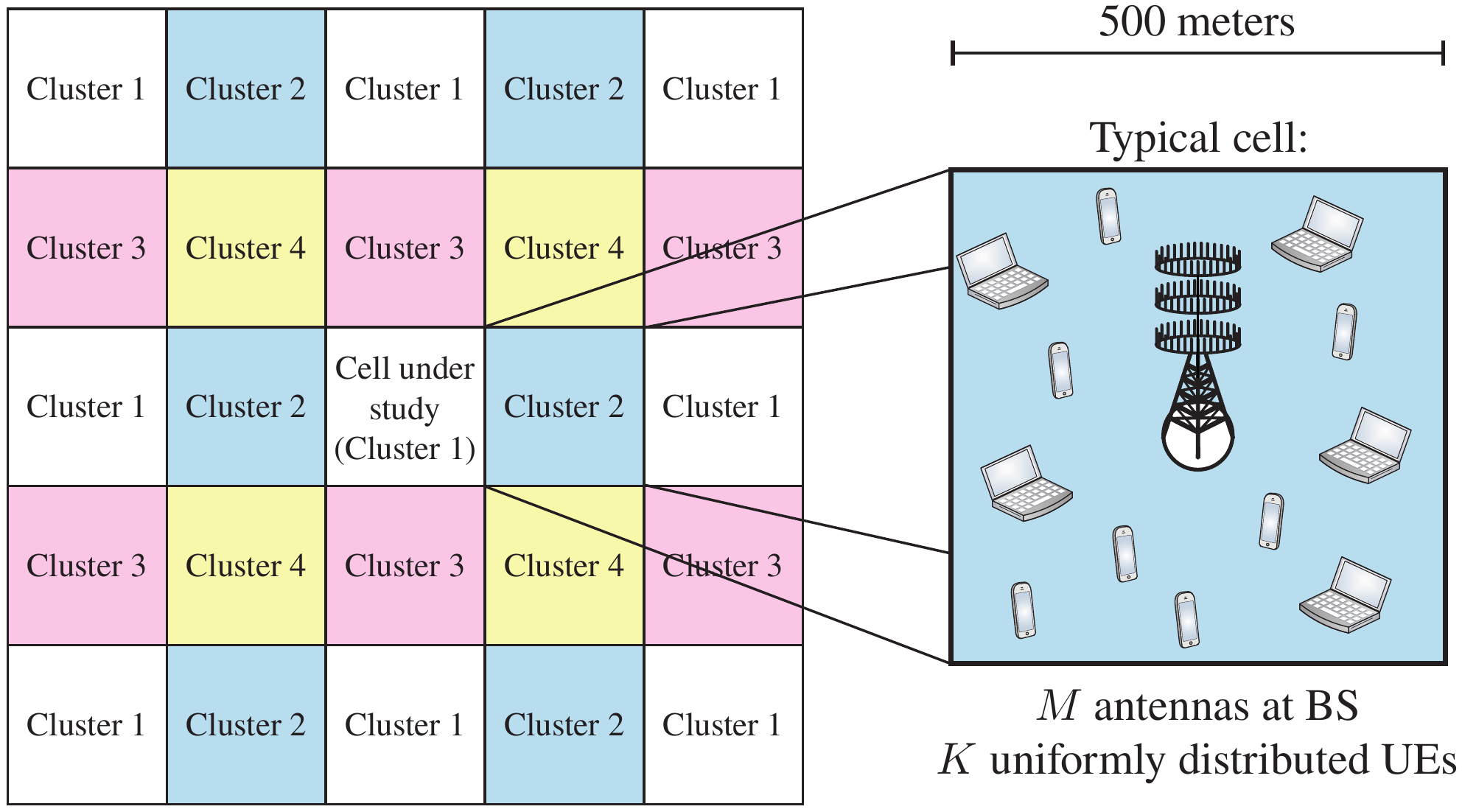}
\end{center}
\vspace{-0.3cm}
\caption{The multi-cell simulation scenario where the cell under study is surrounded by 24 identical cells. The cells are clustered to enable different pilot reuse factors.} \label{multi-cell-scenario} 
\end{figure}

Next, we consider the symmetric multi-cell scenario illustrated in Fig.~\ref{multi-cell-scenario} and concentrate on the cell in the middle. Each cell is a $500 \times 500$ square with uniformly distributed UEs, with the same minimum distance as in the single-cell scenario. We consider only interference that arrives from the two closest cells (in each direction), thus the \emph{cell under study} in Fig.~\ref{multi-cell-scenario} is representative for any cell in the system. Motivated by the single-cell results, we consider only ZF processing and focus on comparing different pilot reuse patterns. As depicted in Fig.~\ref{multi-cell-scenario}, the cells are divided into four clusters. Three different pilot reuse patterns are considered: the same pilots in all cells ($\tau^{(\rm{ul})}=1$), two orthogonal sets of pilots with Cluster 1 and Cluster 4 having the same ($\tau^{(\rm{ul})}=2$), and all clusters have different orthogonal pilots ($\tau^{(\rm{ul})}=4$). Numerical computations of the relative inter-cell interference give $\mathcal{I}_{\rm{PC}} \in \{ 0.5288, \,  0.1163, \, 0.0214 \}$
and $\mathcal{I}_{\rm{PC}^2} \in \{ 0.0405, \,  0.0023, \, 7.82\cdot 10^{-5} \}$, where the values reduce with increasing reuse factor $\tau^{(\rm{ul})}$. Moreover, $\mathcal{I} = 1.5288$ and $\frac{\BW \sigma^2 \PP \mathcal{S}_\mathbf{x}}{\eta} = 1.6022$ in this multi-cell scenario.

\begin{figure}
\begin{center}
\includegraphics[width=\columnwidth]{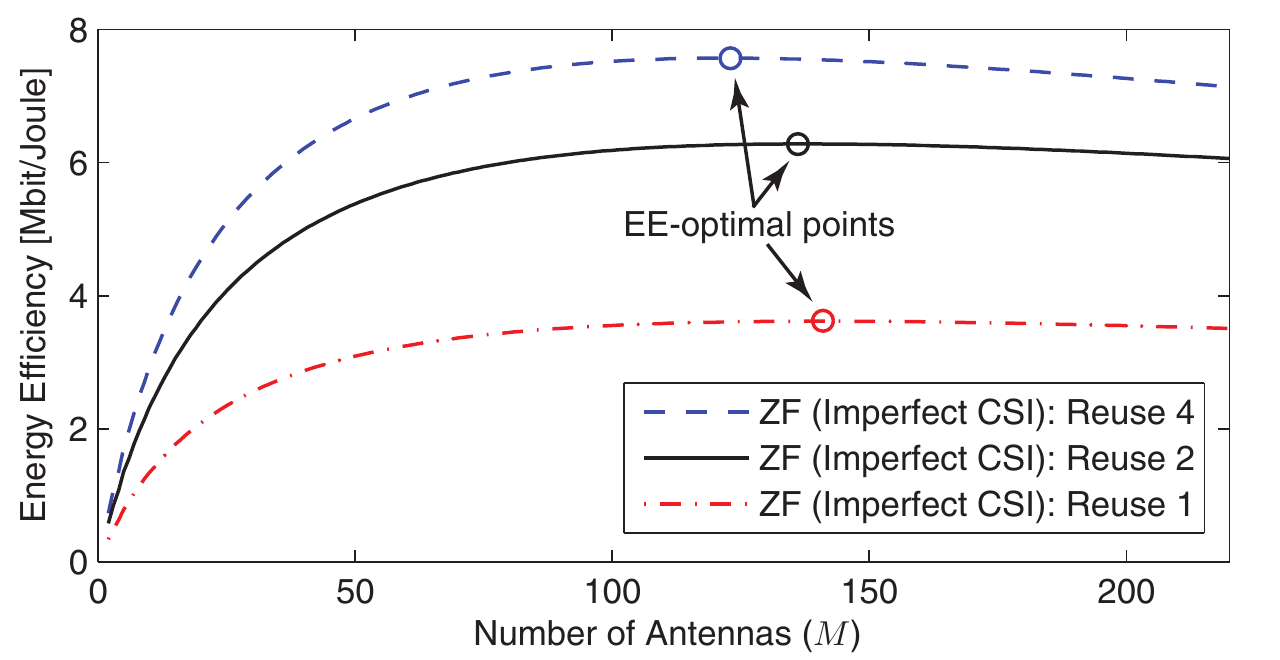}
\end{center}
\vspace{-0.3cm}
\caption{Maximal EE in the multi-cell scenario for different number of BS antennas and different pilot reuse factors.} \label{figure_EE-multicell}
\end{figure}

\begin{figure}
\begin{center}
\includegraphics[width=\columnwidth]{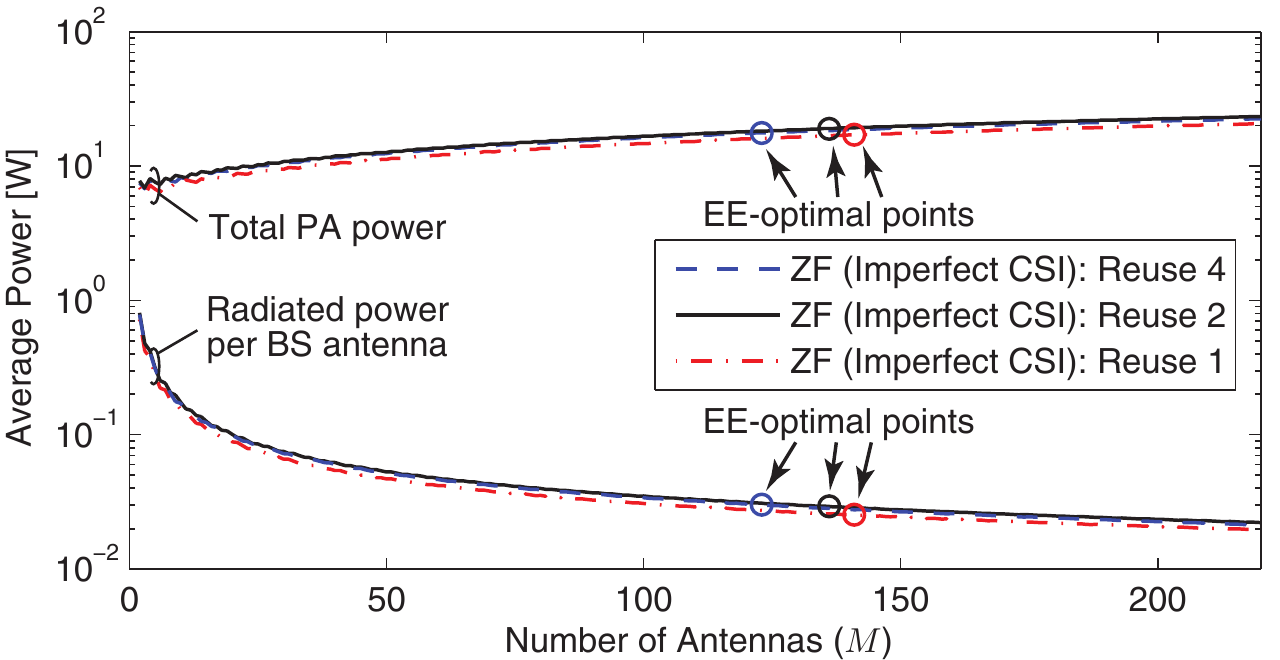}
\end{center}
\vspace{-0.3cm}
\caption{Total PA power at the EE-maximizing solution in the multi-cell scenario, for different number of BS antennas. The radiated power per BS antenna is also shown.} \label{figure_power-multicell}
\end{figure}

\begin{figure}
\begin{center}
\includegraphics[width=\columnwidth]{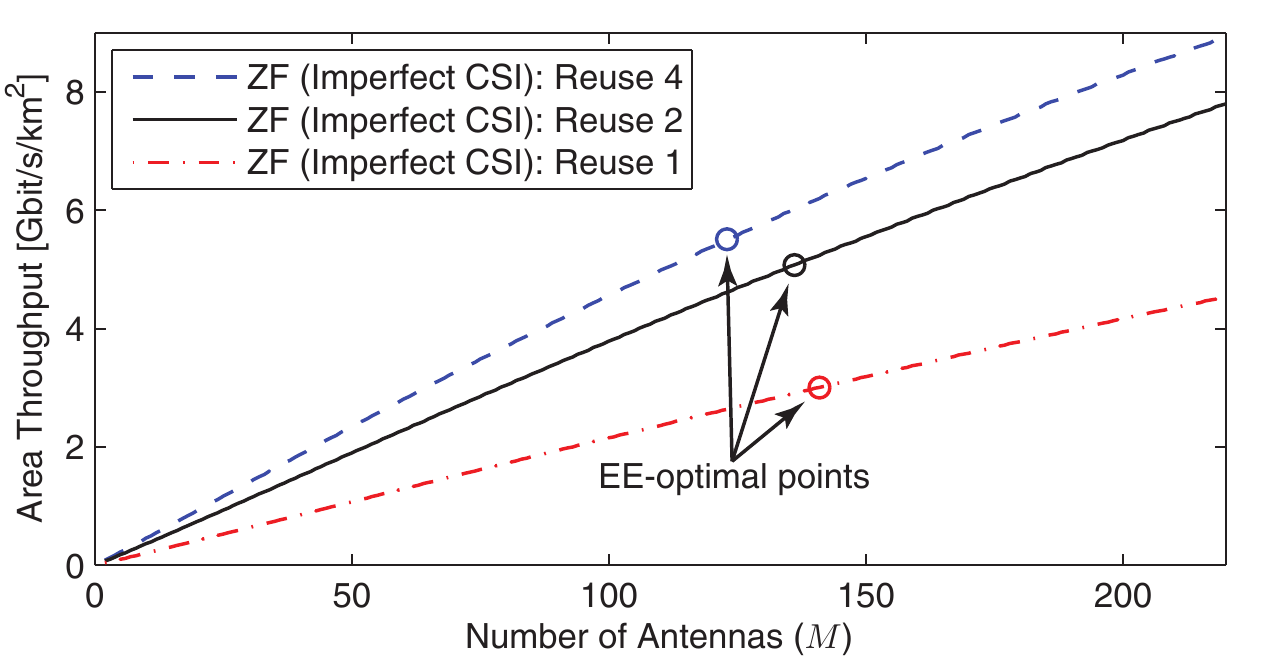}
\end{center}
\vspace{-0.3cm}
\caption{Area throughput at the EE-maximizing solution in the multi-cell scenario, for different number of BS antennas.} \label{figure_rates-multicell}
\end{figure}

The maximal EE for different number of antennas is shown in Fig.~\ref{figure_EE-multicell}, while Fig.~\ref{figure_power-multicell} shows the corresponding PA power (and power per BS antenna) and Fig.~\ref{figure_rates-multicell} shows the area throughput. These figures are very similar to the single-cell counterparts in Figs.~\ref{figure_EE} -- \ref{figure_rates}, but with the main difference that all the numbers are smaller. Hence, the inter-cell interference affects the system by reducing the throughput, reducing the transmit power consumption, and thereby also the EE. Interestingly, the largest pilot reuse factor ($\tau^{(\rm{ul})}=4$) gives the highest EE and area throughput. This shows the necessity of actively mitigating pilot contamination in multi-cell systems. We stress that it is still EE-optimal to increase the transmit power with $M$ (as proved in Corollary \ref{cor:increase-power-with-M} in the single-cell scenario), but at a pace where the power per antenna reduces with $M$.

\begin{figure}
\begin{center}
\includegraphics[width=\columnwidth]{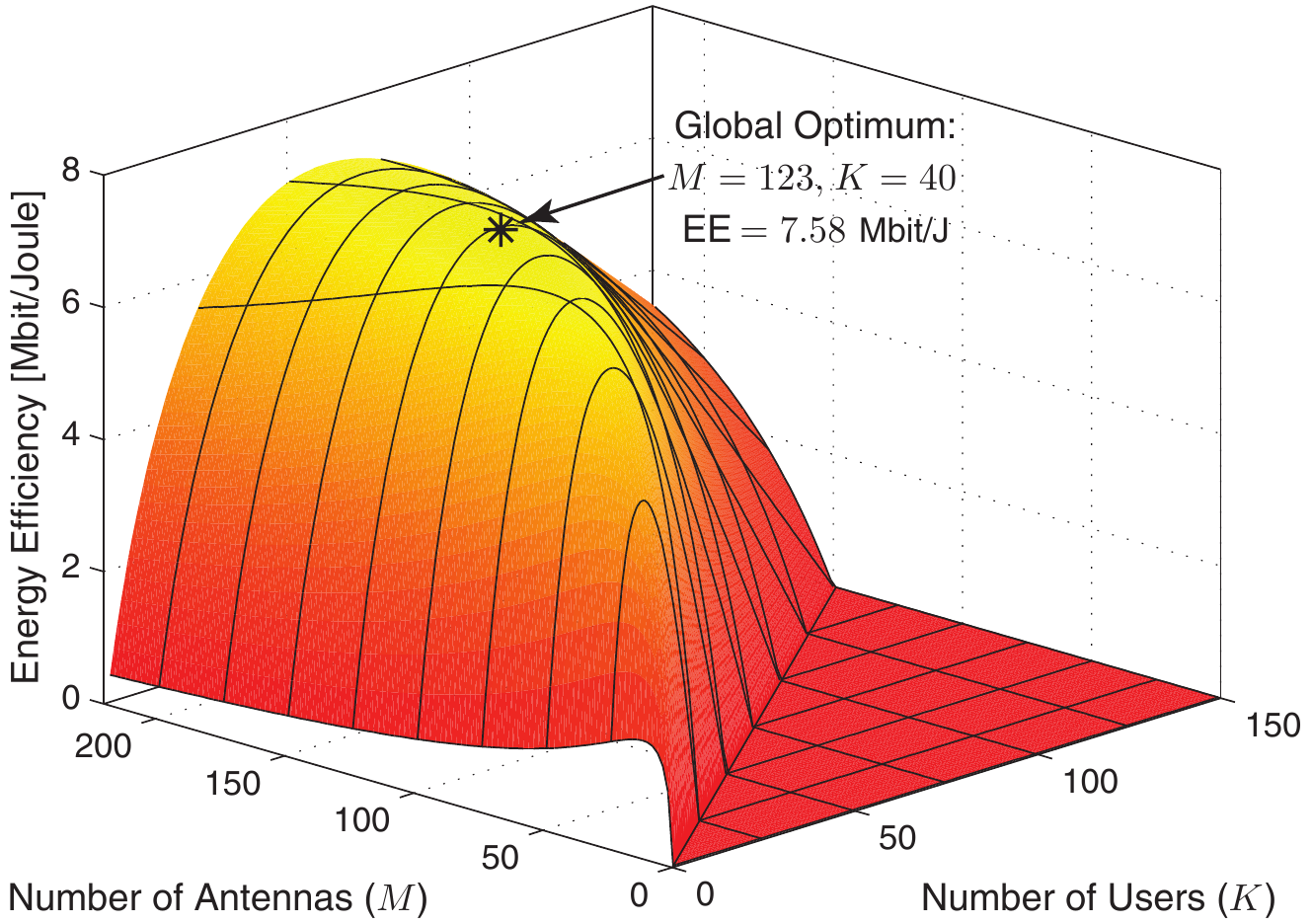}
\end{center}
\vspace{-0.3cm}
\caption{Energy efficiency (in Mbit/Joule) with ZF processing in the multi-cell scenario with pilot reuse 4.} \label{figure_3d_ZFmulticell}
\end{figure}

Finally, the set of achievable EE values is shown in Fig.~\ref{figure_3d_ZFmulticell} for different values of $M$ and $K$. This figure considers a pilot reuse of $\tau^{(\rm{ul})}=4$, since it gives the highest EE. We note that the shape of the set is similar to the single-cell counterpart in Fig.~\ref{figure_3d_ZF}, but the optimal EE value is smaller since it occurs at the smaller system dimensions of $M=123$ and $K=40$ (using a decent spectral efficiency of 1.94 bit/symbol (per UE)). This is mainly due to inter-cell interference, which forces each cell to sacrifice some degrees-of-freedom. We note that the pilot overhead is almost the same as in the single-cell scenario, but the pilot reuse factor gives room for fewer UEs. Nevertheless, we conclude that massive MIMO is the EE-optimal architecture.

\section{Conclusions and Outlook} \label{sec:conclusion}

This paper analyzed how to select the number of BS antennas $M$, number of active UEs $K$, and gross rate $\bar{R}$ (per UE) to maximize the EE in multi-user MIMO systems. Contrary to most prior works, we used a realistic power consumption model that explicitly describes how the total power consumption depends non-linearly on $M$, $K$, and $\bar{R}$. Simple closed-form expressions for the EE-maximizing parameter values and their scaling behaviors were derived under ZF processing with perfect CSI and verified by simulations for other processing schemes, under imperfect CSI, and in symmetric multi-cell scenarios. The applicability in general multi-cell scenarios is an important open problem that we leave for future work.

The EE (in bit/Joule) is a quasi-concave function of $M$ and $K$, thus it has a finite global optimum. Our numerical results show that deploying 100--200 antennas to serve a relatively large number of UEs is the EE-optimal solution using today's circuit technology. We interpret this as massive MIMO setups, but stress that  $M$ and $K$ are at the same order of magnitude (in contrast to the $\frac{M}{K} \gg 1$ assumption in the seminal paper of \cite{Marzetta2010a}). Contrary to common belief, the transmit power should \emph{increase} with $M$ (to compensate for the increasing circuit power) and not decrease. Energy-efficient systems are therefore \emph{not} operating in the low SNR regime, but in a regime where proper interference-suppressing processing (e.g., ZF or MMSE) is highly preferably over interference-ignoring MRT/MRC processing. The radiated power per antenna is, however, decreasing with $M$ and the numerical results show that it is in the range of 10--100 mW. This indicates that massive MIMO can be built using low-power consumer-grade transceiver equipment at the BSs instead of conventional industry-grade high-power equipment.

The analysis was based on spatially uncorrelated fading, while each user might have a unique non-identity channel covariance matrices in practice (e.g., due to limited angular spread and variations in the shadow fading over the array). The statistical information carried in these matrices can be utilized in the scheduler to find statistically compatible users that are likely to interfere less with each other\cite{Huh2012a}. This basically makes the results with imperfect CSI and/or with MRT/MRC processing behave more like ZF processing with perfect CSI does.

The numerical results are stable to small changes in the circuit power coefficients, but can otherwise change drastically. The simulation code is available for download, to enable simple testing of other coefficients. We predict that the circuit power coefficients will decrease over time, implying that the EE-optimal operating point will get a larger value and be achieved using fewer UEs, fewer BS antennas, less transmit power, and more advanced processing.

The system model of this paper assumes that we can serve any number of UEs with any data rate. The problem formulation can be extended to take specific traffic patterns and constraints into account; delay can, for example, be used as an additional dimension to optimize \cite{Kim2009a}. This is outside the scope of this paper, but the closed-form expressions in Theorems \ref{theorem-optimal-K}--\ref{theorem-optimal-rho} can anyway be used to optimize a subset of the parameters while traffic constraints select the others. Another extension is to consider $N$-antenna UEs, where $N>1$. If one stream is sent per UE, one can improve the received signal power proportionally to $N$. If $N$ streams are sent per UE, one can approximate the end performance by treating each UE as $N$ separate UEs in our framework. In both cases, the exact analysis would require a revised and more complicated system model.

\section*{Appendix: Collection of Proofs}

\textbf{Proof of Lemmas \ref{UL-ZF_TX} and \ref{DL-ZF_TX}}: We start by proving Lemma \ref{UL-ZF_TX}. For this purpose,
observe that if a ZF detector is employed, then $\mathbf{D}^{(\rm{ul})}$ in \eqref{D_UL} reduces to a diagonal matrix where the $k$th diagonal entry is $\frac{1}{\PP (M-K) \| \vect{g}_k \|^2}$ (since $|\vect{g}_k^H \vect{h}_k|^2 = 1$ with ZF detection). This implies that
\begin{align} \nonumber
p_k^{(\rm{ul-ZF})} &= \PP(M-K)\sigma^2 \| \vect{g}_k \|^2 \\ &=  \PP(M-K)\sigma^2[(\mathbf{H}^H\mathbf{H})^{-1}]_{k,k}
\end{align}
since $\vect{g}_k$ is the $k$th column of $\mathbf{G}= \mathbf{H} (\mathbf{H}^H\mathbf{H})^{-1}$. Therefore, \eqref{power_UL.11} reduces to
\begin{align}\label{A.1}
P_\mathrm{TX}^{(\rm{ul-ZF})} = \frac{\BW \zeta^{(\rm{ul})}}{\eta^{(\rm{ul})}} \PP(M-K)\sigma^2 \mathbb{E}_{\{\vect{h}_k,\vect{x}_k\}}\left\{ \tr \big( \big(\vect{H}^H \vect{H}\big)^{-1} \big)\right\}
\end{align}
where the expectation is computed with respect to both the channel realizations $\{ \vect{h}_k \}$ and the user locations $\{{\bf{x}}_k\}$. For fixed user locations, we note that $\vect{H}^H \vect{H} \in \mathbb{C}^{K \times K}$ has a complex Wishart distribution with $M$ degrees of freedom and the parameter matrix $\mathbf{\Lambda} = \diag(l(\mathbf{x}_1),l(\mathbf{x}_2),\ldots,l(\mathbf{x}_K))$. By using \cite[Eq.~(50)]{Maiwald2000a}, the inverse first-order moment is
\begin{align} \nonumber
&\mathbb{E}_{\{\vect{h}_k,\vect{x}_k\}}\left\{ \tr \big( \big(\vect{H}^H \vect{H}\big)^{-1} \big)\right\}  \\ &= \mathbb{E}_{\{\vect{x}_k\}}\left\{ \frac{\tr ( \mathbf{\Lambda}^{-1} )}{M-K} \right\}  = \sum_{k=1}^{K} \frac{ \mathbb{E}_{\vect{x}_k} \{ ({ l(\mathbf{x}_k)})^{-1}\} }{M-K}.
\end{align}
Since the expectation with respect to $\mathbf{x}_k$ is the same for all $k$, the average uplink PA power in \eqref{power_UL_ZF} is obtained. The proof of Lemma \ref{DL-ZF_TX} follows the same steps as described above and is omitted for space limitations (we refer to \cite{Bjornson2014b} for details).

\textbf{Proof of Lemma \ref{lemma:optimization-EE}:} We let $\varphi(z) = \frac{g \log(a+bz) }{c + d z + h \log(a+bz)}$ denote the objective function. To prove that this function is
quasi-concave, the level sets $S_{\kappa} =  \{ z \, : \, \varphi(z) \geq \kappa \}$
need to be convex for any $\kappa \in \mathbb{R}$ \cite[Section 3.4]{Boyd2004a}. This set is empty (and thus convex) for $\kappa > \frac{g}{h}$ since $\varphi(z) \leq \frac{g}{h}$. When the set is non-empty, the second-order derivative of $\varphi(z)$ should be negative, which holds for $z > -\frac{a}{b}$ since $ \frac{\partial^2 \varphi(z)}{\partial z^2} = \frac{(h \kappa-g)}{\ln(2)} \frac{b^2}{(a+bz)^2} \leq 0$ for $\kappa \leq \frac{g}{h}$. Hence, $\varphi(z)$ is a quasi-concave function.

If there exists a point $z^\star > -\frac{a}{b}$ such that $\varphi'(z^\star) = 0$, then the quasi-concavity implies that $z^\star$ is the global maximizer and that $\varphi(z)$ is increasing for $z < z^\star$ and decreasing for $z> z^\star$. To prove the existence of $z^\star$, we note that $\varphi'(z) = 0$ if and only if $\frac{1}{\ln(2)}\frac{b  ( c+dz ) }{a+bz}  - d\log(a+bz) =0$ or, equivalently,
\begin{equation} \label{eq:lemma-opt-problem-proof2}
\frac{ bc-ad  }{a+bz} =  d \big( \ln(a+bz) -1 \big).
\end{equation}
Plugging $x = \ln(a+bz) -1$ into \eqref{eq:lemma-opt-problem-proof2} yields 
$\frac{  bc}{de}-\frac{ a  }{e}=x e^x$
whose solution is eventually found to be $x^\star= W(\frac{  bc}{de}-\frac{ a  }{e})$
where $W(\cdot)$ is defined in Definition \ref{def:Lambert}. Finally, we obtain $z^\star = \frac{e^{(x^\star+1)} -a}{b}$.

\vspace{0.2cm}

\textbf{Proof of Theorem \ref{theorem-optimal-K}:} Plugging $ \bar \PP$, $\bar \beta$ and $\bar c$ into \eqref{EE_ZF} leads to the optimization problem
\begin{align} \label{eq:lemma-opt-problem-K}
\maximize{K \, \in \, \mathbb{Z}_{+}} \; \phi(K) 
 \end{align}
 where
 \begin{align}
& \phi(K) = \\
& \frac{K\Big( 1-\frac{\tau_{\mathrm{sum}}  K}{\CU} \Big)   \bar c }{\frac{ \BW \sigma^2 \mathcal{S}_{\mathbf{x}} }{\eta}\bar \PP +\sum\limits_{i=0}^3 \mathcal C_i K^i + \bar \beta \sum\limits_{i=0}^2 \mathcal D_i K^{i+1} + \mathcal A K \Big( 1-\frac{\tau_{\mathrm{sum}} K}{\CU} \Big) \bar c.
} \nonumber
 \end{align}
The function $\phi(K)$ is quasi-concave for $K \in \mathbb{R}$ if the level sets $S_{\kappa} =  \{ K \, : \, \phi(K) \geq \kappa \}$ are convex for any $\kappa \in \mathbb{R}$ \cite[Section 3.4]{Boyd2004a}. This condition is easily verified by differentiation when the coefficients $\mathcal{A}$, $\{\mathcal{C}_i\}$, and $\{\mathcal{D}_i\}$ are non-negative (note that $S_{\kappa}$ is an empty set for $\kappa > \frac{1}{\mathcal A}$). The quasi-concavity implies that the global maximizer of $\phi(K)$ for $K \in \mathbb{R}$ satisfies the stationarity condition $\frac{\partial}{\partial K}\phi(K) = 0$, which is equivalent to finding the roots of the quartic polynomial given in \eqref{eq:polynomial-K}. We denote by $\{K_\ell^{(o)}\}$ the real roots
of \eqref{eq:polynomial-K} and observe that the quasi-concavity of $\phi(K)$ implies that $K^\star$ is either the closest smaller or the closest larger integer.

\vspace{0.2cm}

\textbf{Proof of Corollary \ref{cor:optimal-K}:} This follows from the same line of reasoning used for proving Theorem \ref{theorem-optimal-K}. Observe that if we set $P_\mathrm{CE}=P_{\mathrm{LP}}^{\rm{(ZF)}} = 0$ then $\mathcal C_2 = \mathcal C_3= \mathcal D_1 = \mathcal D_2 = 0$ so that $K^\star$ is obtained as one of the two roots to a quadratic polynomial, for which there are well-known expressions.

\vspace{0.2cm}

\textbf{Proof of Theorem \ref{theorem-optimal-M}:} We need to find the integer value $M^\star \ge K+1$ that maximizes
\begin{equation}\label{eq:1000}
\mathrm{EE}^{(\rm{ZF})} = \frac{ \Big( 1-\frac{\tau_{\mathrm{sum}}  K}{\CU} \Big) \bar R }{ \frac{\BW \sigma^2 \PP \mathcal{S}_\mathbf{x}}{\eta}  + \mathcal C^\prime + M \mathcal D^\prime + \mathcal{A}  \Big( 1-\frac{\tau_{\mathrm{sum}} K}{\CU} \Big) \bar R }.
\end{equation}
where $\mathcal C^\prime $ and $\mathcal D^\prime$ are defined in \eqref{B_prime}. By relaxing $M$ to be real-valued, the maximization of \eqref{eq:1000} is solved by Lemma \ref{lemma:optimization-EE} by setting $a=1-\PP K$, $b = \PP$, $c= { \BW \sigma^2 \mathcal{S}_{\mathbf{x}}} \PP/{\eta}+\mathcal{C}^\prime$, $d = \mathcal D^\prime$, $g=B( 1-{\tau_{\mathrm{sum}} K}/{\CU})$ and $h = \mathcal{A} g$. This lemma proves that $\mathrm{EE} ^{(\rm{ZF})}$ is a quasi-concave function, thus the optimal real-valued solution $M^{(o)}$ in \eqref{eq:lemma-opt-problem-solution} can be transformed into an optimal integer-valued solution as $M^\star=\left\lfloor M^{(o)} \right\rceil$. Finally, we note that the condition $M^\star \geq K+1$ is always satisfied since $\mathrm{EE} ^{(\rm{ZF})}$ is quasi-concave and goes to zero for $M=K$ and when $M \rightarrow \infty$.

\vspace{0.2cm}

\textbf{Proof of Corollary \ref{cor:optimal-M}:} The independence from $\{P_{\mathrm{COD}},P_{\mathrm{DEC}}, P_{\mathrm{BT}}\}$ follows from that $M^\star$  is independent of $\mathcal A$. From Lemma \ref{lemma:lambert-bounds}, we have that the function $e^{W (x) +1}$ is monotonically increasing with $x$. Applying this result to \eqref{eq:optimal_antennas}, it turns out that $M^\star$ is monotonically increasing with $\mathcal{C}^\prime$ and monotonically decreasing with $\mathcal{D}^\prime$. Recalling \eqref{B_prime}, this means that $M^\star$ increases with $\{\mathcal C_{i}\}$ and decreases with $\{\mathcal D_{i}\}$. On the basis of these results, the second part follows from Table \ref{table_coefficients}.

\vspace{0.2cm}

\textbf{Proof of Corollary \ref{cor:optimal-M-scaling}:} The first statement comes from direct application of Lemma \ref{lemma:lambert-bounds} to \eqref{eq:optimal_antennas}, which requires $\frac{  \BW \sigma^2 \mathcal{S}_{\mathbf{x}} }{\eta \mathcal D^\prime}\PP^2 + \frac{\mathcal C^\prime}{\mathcal D^\prime} \PP +  K \PP  - 1   \geq e^2$ (this is satisfied for moderately large values of $\PP$). The scaling law for large values of $\PP$ follows directly from  \eqref{eq:antenna-scaling}.

\vspace{0.2cm}

\textbf{Proof of Theorem \ref{theorem-optimal-rho}:} From \eqref{eq:1000}, the optimal
$\PP$ maximizes
\begin{equation} \label{eq:lemma-opt-problem-rho}
\frac{B\Big( 1-\frac{ \tau_{\mathrm{sum}} K}{\CU} \Big)   \log \Big( 1 + {  \PP(M - K)}
  \Big)
 }{ \frac{\BW \sigma^2 \mathcal{S}_{\mathbf{x}} }{\eta}\PP+\mathcal{C}^\prime + M \mathcal{D}^\prime + \mathcal A \Big( 1-\frac{ \tau_{\mathrm{sum}} K}{\CU} \Big) \log \Big( 1 + {  \PP(M - K)}
  \Big)}
\end{equation}
whose solution follows from Lemma \ref{lemma:optimization-EE} by setting $a=1$, $b = M-K$, $c=\mathcal{C}^\prime + M \mathcal{D}^\prime$, $d=\frac{ \BW \sigma^2 \mathcal{S}_{\mathbf{x}} }{\eta}$, $g=B( 1-\frac{ \tau_{\mathrm{sum}} K}{\CU})$, and $h = \mathcal A g$. The value $\PP^\star$ in \eqref{eq:power-scaling} is always positive since the objective function is quasi-concave and is equal to zero at $\PP=0$ and when $\PP \rightarrow \infty$.

\vspace{0.2cm}

\textbf{Proof of Corollary \ref{cor:increase-power-with-M}:} The lower bound follows from direct application of Lemma \ref{lemma:lambert-bounds} to \eqref{eq:optimal_antennas} under the condition $\frac{\eta (M-K) (\mathcal{C}^\prime + M \mathcal{D}^\prime) }{ \BW \sigma^2 \mathcal{S}_{\mathbf{x}}}- 1 \geq e^2$ (which is satisfied for moderately large values of $M$). The approximation for large $M$ is achieved from  \eqref{power_opt_low_bound} by some simple algebra.

\vspace{0.2cm}

\textbf{Proof of Lemma \ref{lemma:ZF-rate-imperfect}:} Let the uplink pilot power of the $k$th UE be $\frac{\PP \sigma^2}{l(\vect{x}_k)}$ and consider the use of orthogonal pilot sequences of length $K \tau^{(\rm{ul})}$. By using MMSE estimation \cite{Bjornson2010a}, we obtain a channel estimate $\hat{\vect{h}}_k \sim \mathcal{CN} \big( \vect{0}_N, \frac{l(\vect{x}_k)}{1+\frac{1}{\PP K \tau^{(\rm{ul})}} } \vect{I}_k \big)$ with the estimation error covariance matrix 
\begin{equation}
l(\vect{x}_k) \big( 1- \frac{1}{1+\frac{1}{\PP K \tau^{(\rm{ul})}} } \big) \vect{I}_N. 
\end{equation}
We apply approximate ZF in the uplink and downlink by treating the channel estimates as the true channels. By treating the estimation errors as noise with a variance that is averaged over the channel realizations, the $k$th UE achieves the average gross rate
\begin{equation}
\bar{R} = \BW \log \Bigg( 1+ \frac{p_k^{(\rm{ul})}}{\| \vect{g}_k\|^2 \big( \sigma^2 + \big(1- \frac{1}{1+\frac{1}{\PP K \tau^{(\rm{ul})}} } \big) K \PP \sigma^2 \big)} \Bigg)
\end{equation}
which is equivalent to  \eqref{eq:gross-rate-imperfect} for the uplink transmit powers $p_k^{(\rm{ul})} = \frac{\PP \sigma^2 (M-K) \| \vect{g}_k\|^2 }{ 1+\frac{1}{\PP K \tau^{(\rm{ul})}}}$. The downlink rate is derived analogously and it is straightforward to compute the average total PA power.

\vspace{0.2cm}

\textbf{Proof of Lemma \ref{lemma:ZF-rate-multicell}:} We assume that the uplink power for UE $k$ in cell $j$ is
\begin{equation}
p_{jk}^{(\rm{ul})} = \frac{\sigma^2 \PP(M-K) \| \vect{g}_{jk}\|^2 }{\Big(1+\mathcal{I}_{\rm{PC}}+\frac{1}{\PP K \tau^{(\rm{ul})}}\Big)}
\end{equation}
during data transmission, where $\vect{g}_{jk}$ is the receive filter. Under approximate ZF we have
\begin{equation}
\mathbb{E}\{ \| \vect{g}_{jk} \|^2 \} = \frac{1+\mathcal{I}_{\rm{PC}}+\frac{1}{\PP K \tau^{(\rm{ul})}}}{(M-K) l_j(\vect{x}_{jk}) }
\end{equation}
when averaging over the channel realizations, thus the average UE power is the same as in Lemma \ref{UL-ZF_TX}. The channel-averaged value $p_{jk}^{(\rm{ul-pilot})} = \frac{\sigma^2 \PP }{l_j(\vect{x}_{jk})}$ is used for pilot transmission, since it can only depend on channel statistics. If the BS applies MMSE estimation \cite{Bjornson2010a} and is unaware of the UE positions in other cells, the average interference from cells with orthogonal pilots is $\| \vect{g}_{jk}\|^2 \PP K \tau^{(\rm{ul})} \sum_{\ell \not \in \mathcal{Q}_j} \mathcal{I}_{j \ell}$. The average interference from the cells using the same pilots is
\begin{align} \nonumber
& \PP (M-K) \| \vect{g}_{jk}\|^2 \frac{\mathcal{I}^{(\rm{PC})}}{1+\mathcal{I}^{(\rm{PC})}+\frac{1}{\PP K \tau^{(\rm{ul})}}} \\ &+ \| \vect{g}_{jk}\|^2 \PP K \tau^{(\rm{ul})}    \Big( \sum_{\ell \in \mathcal{Q}_j}\mathcal{I}_{j \ell} - \frac{ \sum_{\ell \in \mathcal{Q}_j} \mathcal{I}_{j \ell}^2}{1+\mathcal{I}^{(\rm{PC})}+\frac{1}{\PP K \tau^{(\rm{ul})}}}\Big )
\end{align}
where the first term is due to PC and the second is due to channel uncertainty. Putting this together, we achieve the gross rate in \eqref{eq:gross-rate-multicell} in the uplink. The same expression is achieved in the downlink by treating channel uncertainty as noise and exploiting the cell symmetry.

\bibliographystyle{IEEEtran}
\bibliography{IEEEabrv,refs}

\begin{thebibliography}{10}
\providecommand{\url}[1]{#1}
\csname url@samestyle\endcsname
\providecommand{\newblock}{\relax}
\providecommand{\bibinfo}[2]{#2}
\providecommand{\BIBentrySTDinterwordspacing}{\spaceskip=0pt\relax}
\providecommand{\BIBentryALTinterwordstretchfactor}{4}
\providecommand{\BIBentryALTinterwordspacing}{\spaceskip=\fontdimen2\font plus
\BIBentryALTinterwordstretchfactor\fontdimen3\font minus
  \fontdimen4\font\relax}
\providecommand{\BIBforeignlanguage}[2]{{%
\expandafter\ifx\csname l@#1\endcsname\relax
\typeout{** WARNING: IEEEtran.bst: No hyphenation pattern has been}%
\typeout{** loaded for the language `#1'. Using the pattern for}%
\typeout{** the default language instead.}%
\else
\language=\csname l@#1\endcsname
\fi
#2}}
\providecommand{\BIBdecl}{\relax}
\BIBdecl

\bibitem{EARTH_D23}
\BIBentryALTinterwordspacing
G.~Auer and et~al., \emph{D2.3: Energy efficiency analysis of the reference
  systems, areas of improvements and target breakdown}.\hskip 1em plus 0.5em
  minus 0.4em\relax INFSO-ICT-247733 EARTH, ver.~2.0, 2012. [Online].
  Available: \url{http://www.ict-earth.eu/}
\BIBentrySTDinterwordspacing

\bibitem{Chen2011a}
Y.~Chen, S.~Zhang, S.~Xu, and G.~Li, ``Fundamental trade-offs on green wireless
  networks,'' \emph{{IEEE} Commun. Mag.}, vol.~49, no.~6, pp. 30--37, 2011.

\bibitem{Smart_2020}
``Smart 2020: Enabling the low carbon economy in the information age,'' The
  Climate Group and Global e-Sustainability Initiative (GeSI), Tech. Rep.,
  2008.

\bibitem{GreenTouch}
\BIBentryALTinterwordspacing
Green Touch Consortium, Tech. Rep. [Online]. Available:
  \url{http://www.greentouch.org}
\BIBentrySTDinterwordspacing

\bibitem{Tombaz2011a}
S.~Tombaz, A.~V\"{a}stberg, and J.~Zander, ``Energy- and cost-efficient
  ultra-high-capacity wireless access,'' \emph{{IEEE} Wireless Commun. Mag.},
  vol.~18, no.~5, pp. 18--24, 2011.

\bibitem{Bjornson2014a}
E.~Bj{\"{o}}rnson, J.~Hoydis, M.~Kountouris, and M.~Debbah, ``Massive {MIMO}
  systems with non-ideal hardware: Energy efficiency, estimation, and capacity
  limits,'' \emph{{IEEE} Trans. Inf. Theory}, vol.~60, no.~11, pp. 7112--7139,
  2014.

\bibitem{Marzetta2010a}
T.~Marzetta, ``Noncooperative cellular wireless with unlimited numbers of base
  station antennas,'' \emph{{IEEE} Trans. Wireless Commun.}, vol.~9, no.~11,
  pp. 3590--3600, 2010.

\bibitem{Rusek2013a}
F.~Rusek, D.~Persson, B.~Lau, E.~Larsson, T.~Marzetta, O.~Edfors, and
  F.~Tufvesson, ``Scaling up {MIMO}: Opportunities and challenges with very
  large arrays,'' \emph{{IEEE} Signal Process. Mag.}, vol.~30, no.~1, pp.
  40--60, 2013.

\bibitem{Hoydis2013a}
J.~Hoydis, S.~ten Brink, and M.~Debbah, ``Massive {MIMO} in the {UL/DL} of
  cellular networks: How many antennas do we need?'' \emph{{IEEE} J. Sel. Areas
  Commun.}, vol.~31, no.~2, pp. 160--171, 2013.

\bibitem{Ngo2013a}
H.~Ngo, E.~Larsson, and T.~Marzetta, ``Energy and spectral efficiency of very
  large multiuser {MIMO} systems,'' \emph{{IEEE} Trans. Commun.}, vol.~61,
  no.~4, pp. 1436--1449, 2013.

\bibitem{Miao2013a}
G.~Miao, ``Energy-efficient uplink multi-user {MIMO},'' \emph{{IEEE} Trans.
  Wireless Commun.}, vol.~12, no.~5, pp. 2302--2313, 2013.

\bibitem{Hu2014a}
Y.~Hu, B.~Ji, Y.~Huang, F.~Yu, and L.~Yang, ``Energy-efficiency resource
  allocation of very large multi-user {MIMO} systems,'' \emph{Wireless Netw.},
  2014.

\bibitem{Bjornson2013e}
E.~Bj{\"{o}}rnson, M.~Kountouris, and M.~Debbah, ``Massive {MIMO} and small
  cells: Improving energy efficiency by optimal soft-cell coordination,'' in
  \emph{Proc.~Int.~Conf.~Telecommun.~(ICT)}, 2013.

\bibitem{Ha2013a}
D.~Ha, K.~Lee, and J.~Kang, ``Energy efficiency analysis with circuit power
  consumption in massive {MIMO} systems,'' in \emph{Proc.~IEEE
  Int.~Symp.~Personal, Indoor and Mobile Radio Commun.~(PIMRC)}, 2013.

\bibitem{Yang2013a}
H.~Yang and T.~Marzetta, ``Total energy efficiency of cellular large scale
  antenna system multiple access mobile networks,'' in \emph{Proc. IEEE Online
  Conference on Green Communications (OnlineGreenComm)}, 2013.

\bibitem{Mohammed2014a}
\BIBentryALTinterwordspacing
S.~Mohammed, ``Impact of transceiver power consumption on the energy efficiency
  spectral efficiency tradeoff of zero-forcing detector in massive {MIMO}
  systems,'' 2014, submitted. [Online]. Available:
  \url{http://arxiv.org/abs/1401.4907v1}
\BIBentrySTDinterwordspacing

\bibitem{Bjornson2013d}
E.~Bj{\"{o}}rnson and E.~Jorswieck, ``Optimal resource allocation in
  coordinated multi-cell systems,'' \emph{Foundations and Trends in
  Communications and Information Theory}, vol.~9, no. 2-3, pp. 113--381, 2013.

\bibitem{Zetterberg2011a}
P.~Zetterberg, ``Experimental investigation of {TDD} reciprocity-based
  zero-forcing transmit precoding,'' \emph{EURASIP J. on Adv. in Signal
  Process.}, Jan. 2011.

\bibitem{Shepard2012a}
C.~Shepard, H.~Yu, N.~Anand, L.~Li, T.~Marzetta, R.~Yang, and L.~Zhong,
  ``Argos: Practical many-antenna base stations,'' in \emph{Proc.~ACM MobiCom},
  2012.

\bibitem{LTE2010b}
\emph{Further advancements for {E-UTRA} physical layer aspects (Release
  9)}.\hskip 1em plus 0.5em minus 0.4em\relax {3GPP} {TS} 36.814, Mar. 2010.

\bibitem{Gao2015a}
\BIBentryALTinterwordspacing
X.~Gao, O.~Edfors, F.~Rusek, and F.~Tufvesson, ``Massive {MIMO} in real
  propagation environments,'' \emph{{IEEE} Trans. Wireless Commun.}, 2014,
  submitted. [Online]. Available: \url{http://arxiv.org/abs/1403.3376}
\BIBentrySTDinterwordspacing

\bibitem{Pillai2005a}
S.~Pillai, T.~Suel, and S.~Cha, ``The {Perron}-{Frobenius} theorem: some of its
  applications,'' \emph{{IEEE} Signal Process. Mag.}, vol.~22, no.~2, pp.
  62--75, 2005.

\bibitem{Wiesel2006a}
A.~Wiesel, Y.~Eldar, and S.~Shamai, ``Linear precoding via conic optimization
  for fixed {MIMO} receivers,'' \emph{{IEEE} Trans. Signal Process.}, vol.~54,
  no.~1, pp. 161--176, 2006.

\bibitem{Boche2002a}
H.~Boche and M.~Schubert, ``A general duality theory for uplink and downlink
  beamforming,'' in \emph{Proc.~IEEE VTC-Fall}, 2002, pp. 87--91.

\bibitem{Caire2013a}
\BIBentryALTinterwordspacing
Y.-G. Lim, C.-B. Chae, and G.~Caire, ``Performance analysis of massive {MIMO}
  for cell-boundary users,'' \emph{{IEEE} J. Sel. Topics Signal Process.},
  2013, submitted. [Online]. Available: \url{http://arxiv.org/abs/1309.7817}
\BIBentrySTDinterwordspacing

\bibitem{Cui2004a}
S.~Cui, A.~Goldsmith, and A.~Bahai, ``Energy-efficiency of {MIMO} and
  cooperative {MIMO} techniques in sensor networks,'' \emph{{IEEE} J. Sel.
  Areas Commun.}, vol.~22, no.~6, pp. 1089--1098, 2004.

\bibitem{Mezghani2011a}
A.~Mezghani and J.~A. Nossek, ``Power efficiency in communication systems from
  a circuit perspective,'' in \emph{Proc. IEEE Int. Symp. Circuits and Systems
  (ISCAS)}, 2011, pp. 1896--1899.

\bibitem{Kumar2011a}
R.~Kumar and J.~Gurugubelli, ``How green the {LTE} technology can be?'' in
  \emph{Proc.~Wireless VITAE}, 2011.

\bibitem{Boyd2008a}
\BIBentryALTinterwordspacing
S.~Boyd and L.~Vandenberghe, ``Numerical~linear~algebra~background.'' [Online].
  Available: \url{www.ee.ucla.edu/ee236b/lectures/num-lin-alg.pdf}
\BIBentrySTDinterwordspacing

\bibitem{Mukherjee2014a}
\BIBentryALTinterwordspacing
S.~Mukherjee and S.~Mohammed, ``On the energy-spectral efficiency trade-off of
  the {MRC} receiver in massive {MIMO} systems with transceiver power
  consumption,'' 2014. [Online]. Available:
  \url{http://arxiv.org/abs/1404.3010}
\BIBentrySTDinterwordspacing

\bibitem{Hoorfar2008a}
A.~Hoorfar and M.~Hassani, ``Inequalities on the {Lambert $W$} function and
  hyperpower function,'' \emph{{J. Inequalities in Pure and Applied Math.}},
  vol.~9, no.~2, pp. 1--5, 2008.

\bibitem{Shmakov2011a}
F.~Shmakov, ``A universal method of solving quartic equations,'' \emph{Int. J.
  Pure and Applied Math.}, vol.~71, no.~2, pp. 251--259, 2011.

\bibitem{Kang2011a}
D.~Kang, D.~Kim, Y.~Cho, J.~Kim, B.~Park, C.~Zhao, and B.~Kim, ``1.6--2.1 {GHz}
  broadband doherty power amplifiers for {LTE} handset applications,'' in
  \emph{Proc.~IEEE MTT-S Int. Microwave Symp. Digest}, 2011.

\bibitem{Tombaz2012a}
S.~Tombaz, K.~Sung, and J.~Zander, ``Impact of densification on energy
  efficiency in wireless access networks,'' in \emph{Proc.~IEEE Global
  Commun.~Conf.~(GLOBECOM)}, 2012.

\bibitem{Parker2009a}
M.~Parker, ``High-performance floating-point implementation using {FPGAs},'' in
  \emph{Proc.~IEEE MILCOM}, 2009.

\bibitem{Bjornson2014b}
E.~Bj{\"{o}}rnson, L.~Sanguinetti, J.~Hoydis, and M.~Debbah, ``Designing
  multi-user {MIMO} for energy efficiency: When is massive {MIMO} the answer?''
  in \emph{Proc. IEEE Wireless Commun. and Networking Conf. (WCNC)}, 2014.

\bibitem{Huh2012a}
H.~Huh, G.~Caire, H.~Papadopoulos, and S.~Ramprashad, ``Achieving ``massive
  {MIMO}'' spectral efficiency with a not-so-large number of antennas,''
  \emph{{IEEE} Trans. Wireless Commun.}, vol.~11, no.~9, pp. 3226--3239, 2012.

\bibitem{Kim2009a}
H.~Kim, C.-B. Chae, G.~de~Veciana, and R.~Heath, ``A cross-layer approach to
  energy efficiency for adaptive {MIMO} systems exploiting spare capacity,''
  \emph{{IEEE} Trans. Wireless Commun.}, vol.~8, no.~8, pp. 4264--4275, 2009.

\bibitem{Maiwald2000a}
D.~Maiwald and D.~Kraus, ``Calculation of moments of complex {Wishart} and
  complex inverse {Wishart} distributed matrices,'' \emph{{IEE} Proc.~Radar
  Sonar Navig.}, vol. 147, no.~4, pp. 162--168, 2000.

\bibitem{Boyd2004a}
S.~Boyd and L.~Vandenberghe, \emph{Convex Optimization}.\hskip 1em plus 0.5em
  minus 0.4em\relax Cambridge University Press, 2004.

\bibitem{Bjornson2010a}
E.~Bj{\"{o}}rnson and B.~Ottersten, ``A framework for training-based estimation
  in arbitrarily correlated {Rician} {MIMO} channels with {Rician}
  disturbance,'' \emph{{IEEE} Trans. Signal Process.}, vol.~58, no.~3, pp.
  1807--1820, 2010.

\end{thebibliography}

\end{document}